\documentclass[preprint]{aastex}
\usepackage[usenames,dvipsnames]{color}
\usepackage{natbib,graphicx,latexsym,lscape,pdflscape,url,pdfpages}
\pdfoutput=1

\newcommand{\Hipparcos}{{\sl Hipparcos}}
\newcommand{\HST}{{\sl HST}}

\newcommand{\Msun}{\mbox{$M_{\sun}$}}

\newcommand{\Lsun}{\mbox{$L_{\sun}$}}

\newcommand{\Mjup}{\mbox{$M_{\rm Jup}$}}

\newcommand{\degree}{\mbox{$^{\circ}$}}
\newcommand{\perpix}{\mbox{pixel$^{-1}$}}

\newcommand{\kms}{\mbox{km\,s$^{-1}$}}

\newcommand{\Ks}{\mbox{$K_S$}}
\newcommand{\Ynirc}{\mbox{$Y_{\rm NIRC2}$}}

\newcommand{\Mtot}{\mbox{$M_{\rm tot}$}}

\newcommand{\Lbol}{\mbox{$L_{\rm bol}$}}

\newcommand{\Teff}{\mbox{$T_{\rm eff}$}}
\newcommand{\logg}{\mbox{$\log(g)$}}

\newcommand{\Lp}{\mbox{${L^\prime}$}}

\newcommand{\logRHK}{\mbox{${\log{R^\prime_{\rm HK}}}$}}
\newcommand{\logRX}{\mbox{${\log{R_{\rm X}}}$}}
\newcommand{\logtyr}{\mbox{${\log(t/{\rm yr})}$}}
\newcommand{\tauc}{\mbox{${\tau_{\rm conv}}$}}

\newcommand{\glA}{Gl~417A}
\newcommand{\glB}{Gl~417B}
\newcommand{\glC}{Gl~417C}
\newcommand{\glBC}{Gl~417BC}

\newcommand{\hdA}{HD~130948A}
\newcommand{\hdB}{HD~130948B}
\newcommand{\hdC}{HD~130948C}
\newcommand{\hdBC}{HD~130948BC}

\shorttitle{A Substellar Luminosity Problem?}
\shortauthors{Dupuy, Liu, \& Ireland}

\begin{document}

\title{New Evidence for a Substellar Luminosity Problem: \\ 
  Dynamical Mass for the Brown Dwarf Binary Gl~417BC\altaffilmark{*}}

\author{Trent J.\ Dupuy,\altaffilmark{1,2}
        Michael C.\ Liu,\altaffilmark{3} and
        Michael J.\ Ireland\altaffilmark{4,5,6}}

      \altaffiltext{*}{Data presented herein were obtained at the
        W.M.\ Keck Observatory, which is operated as a scientific
        partnership among the California Institute of Technology, the
        University of California, and the National Aeronautics and
        Space Administration. The Observatory was made possible by the
        generous financial support of the W.M.\ Keck Foundation.}

      \altaffiltext{1}{The University of Texas at Austin, Department
        of Astronomy, 2515 Speedway C1400, Austin, TX 78712, USA}

      \altaffiltext{2}{Harvard-Smithsonian Center for Astrophysics,
        60 Garden Street, Cambridge, MA 02138}

      \altaffiltext{3}{Institute for Astronomy, University of Hawai`i,  
        2680 Woodlawn Drive, Honolulu, HI 96822, USA}

      \altaffiltext{4}{Department of Physics and Astronomy, Macquarie
        University, NSW 2109, Australia}

      \altaffiltext{5}{Research School of Astronomy \& Astrophysics,
        Australian National University, Canberra ACT 2611, Australia}

      \altaffiltext{6}{Australian Astronomical Observatory, PO Box
        296, Epping, NSW 1710, Australia}

\begin{abstract}

  We present new evidence for a problem with cooling rates predicted
  by substellar evolutionary models that implies model-derived masses
  in the literature for brown dwarfs and directly imaged planets may
  be too high.  Based on our dynamical mass for Gl~417BC (L4.5+L6) and
  a gyrochronology system age from its young, solar-type host star,
  commonly used models predict luminosities 0.2--0.4\,dex lower than
  we observe.  This corroborates a similar luminosity--age discrepancy
  identified in our previous work on the L4+L4 binary HD~130948BC,
  which coincidentally has nearly identical component masses
  ($\approx$50--55\,\Mjup) and age ($\approx$800\,Myr) as Gl~417BC.
  Such a luminosity offset would cause systematic errors of 15\%--25\%
  in model-derived masses at this age. After comparing different
  models, including cloudless models that should not be appropriate
  for mid-L dwarfs like Gl~417BC and HD~130948BC but actually match
  their luminosities better, we speculate the observed over-luminosity
  could be caused by opacity holes (i.e., patchy clouds) in these
  objects.  Moreover, from hybrid substellar evolutionary models that
  account for cloud disappearance we infer the corresponding phase of
  over-luminosity may extend from a few hundred Myr up to a few Gyr
  and cause masses to to be over-estimated by up to 25\%, even well
  after clouds disappear from view entirely. Thus, the range of of
  ages and spectral types affected by this potential systematic shift
  in luminosity evolution would encompass most known directly imaged
  gas-giants and field brown dwarfs.

\end{abstract}


%----------------------------------------------------------------------%

\section{Introduction}

Models of substellar evolution have been notoriously under constrained
in the brown dwarf regime, but the last several years has seen
significant progress.  An increasing number of brown dwarf visual
binaries have dynamical masses \citep{2001ApJ...560..390L,
  2008ApJ...689..436L, 2009ApJ...692..729D, 2009ApJ...699..168D,
  2010ApJ...721.1725D, 2010ApJ...711.1087K}, and there are now several
transiting brown dwarfs that provide tests of the mass--radius
relationship \citep[e.g.,][]{2006Natur.440..311S, 2008A&A...491..889D,
  2011ApJ...726L..19A, 2011ApJ...730...79J, 2012ApJ...761..123S}.
However, the total energy output of substellar objects as a function
of mass and age has still barely been tested, despite the fact that
these fundamental predictions underpin the mass estimates for all
brown dwarfs and extrasolar planets that lack directly determined
masses.

Until now there has been only been one system that enables a robust
test of substellar luminosity evolution because of the demanding
requirements of both precise mass, age, and luminosity
determinations.\footnote{In principle, the giant planets of the solar
  system also allow tests of substellar luminosity evolution but with
  very different assumptions for internal composition and structure
  such as the possible presence of metal-rich cores.  In the latest
  models of \citet{2011ApJ...729...32F}, Jupiter is only slightly
  (0.04\,dex) under-luminous compared to models, while Saturn is
  0.20\,dex over-luminous.  Special mechanisms have been proposed to
  explain Saturn's excess luminosity such as helium rain
  \citep[e.g.,][]{1980Sci...208..746S} or, more recently, layered
  convection driven by a steep molecular weight gradient
  \citep{2013NatGe...6..347L}.}  In \citet{2009ApJ...692..729D}, we
found that the components of the brown dwarf visual binary \hdBC\ were
$\approx$2$\times$ more luminous than expected from models given their
age and mass. This was a surprising result, given that the bulk
properties of brown dwarfs from evolutionary models were thought to be
relatively robust against the boundary conditions of models by several
hundred Myr.  Furthermore, there has been no satisfactory theoretical
explanation of how such a luminosity problem might arise; for example,
even custom magnetic models of \hdBC\ cannot match the observations
\citep{2010ApJ...713.1249M}.  We also note that an earlier dynamical
mass measurement for the substellar companion Gl~802B hinted at a
similar luminosity problem due to its likely thick disk membership
being inconsistent with an age of $\sim$2\,Gyr inferred from
evolutionary models using its mass and near-infrared flux
\citep{2008ApJ...678..463I}.  With no other systems to test models it
has not yet been clear if there truly is a problem in predictions of
substellar luminosity evolution, which could affect model-derived mass
estimates and thereby have wide-ranging implications, or if our one
test case was simply an unfortunate outlier.

We present a dynamical mass for the L~dwarf binary \glBC\ that
provides new evidence for the same substellar luminosity problem found
for \hdBC.  \citet{2000AJ....120..447K} originally identified the
Gl~417BC system (unresolved) as the lithium-bearing L4.5 dwarf
2MASSW~J1112257+354813.  \citet{2001AJ....121.3235K} subsequently
found that it was co-moving at a distance of 90$\arcsec$ from the star
Gl~417, now dubbed \glA.\footnote{Some other names for \glA\ are
  HD~97334, BD+36~2162, MN~UMa, HR~4345, and HIP~54745.} The revised
\Hipparcos\ parallax for \glA\ is $45.61\pm0.44$\,mas
\citep{2007A&A...474..653V}, giving a distance of $21.93\pm0.21$\,pc
to the system and projected separation of $1970\pm20$\,AU for the
Gl~417AB pair.  \citet{2001AJ....121.3235K} used age indicators for
the primary star such as chromospheric activity, rotation, and lithium
absorption to estimate an age of 80--300\,Myr for the system. This age
range implied a substellar mass for 2MASSW~J1112257+354813,
corroborated by the detection of lithium in its spectrum.
\citet{2003AJ....126.1526B} presented \textsl{Hubble Space Telescope}
Wide-Field Planetary Camera~2 (\HST/WFPC2) images resolving this
object as a binary, named \glBC, with a projected separation of
70\,mas (or 1.5\,AU) and an estimated orbital period of
$<$10~years. This made \glBC\ one of the most likely substellar
binaries to yield a dynamical mass from determining the visual orbit
of the brown dwarfs around each other. However, unlike some substellar
companions, \glBC\ is too distant from its host star to use natural
guide star adaptive optics (AO), and there has been no resolved
astrometry of \glBC\ published in more than a decade since it was
discovered to be a binary.

We have obtained Keck laser guide star (LGS) AO imaging of \glBC.
Combined with the original \HST\ imaging, our astrometric data set
spans more than 13~years and enables us to determine a precise total
dynamical mass for this binary. Moreover, \glA\ provides a much more
precise age estimate than is typically possible for stars in the field
population because of its youth.  Therefore, \glBC\ now joins \hdBC\
as the only brown dwarf binaries with precisely determined masses
\emph{and} ages.

%----------------------------------------------------------------------%

\section{Astrometric Monitoring of \glBC \label{sec:astrom}}

\subsection{Keck/NIRC2 LGS AO \label{sec:keck}}

We used the facility near-infrared camera NIRC2 with the LGS AO system
at the Keck~II telescope \citep{2006PASP..118..297W,
  2006PASP..118..310V} to image the binary \glBC\ at eight epochs
spanning 2007~March~25~UT to 2014~May~9~UT.  We obtained data in
standard Mauna Kea Observatories (MKO) photometric bandpasses
\citep{mkofilters1, mkofilters2}.\footnote{See the appendix of
  \citet{2012ApJ...758...57L} for a discussion of the $Y$ bandpass of
  NIRC2 compared to other photometric systems.}  The LGS was kept
centered in NIRC2's narrow camera field-of-view while we obtained
dithered images of the target.  The wavefront sensor recorded flux
from the LGS equivalent to a $V \approx 9.5$--10.4~\,mag star.  For
tip-tilt correction we used the star SDSS~J111229.47+354813.2, which
is 46$\arcsec$ away from \glBC, and the lower bandwidth sensor
monitoring this source recorded flux equivalent to a $R \approx
17.5$--18.0\,mag star.  We note that this tip-tilt star was not
provided by the standard Keck IDL routine \texttt{FINDTTREF}---the
tool used by most Keck LGS observers to determine if a target has
suitable reference stars---because it does not exist in the USNO-B
catalog \citep{2003AJ....125..984M}.  However, we noticed it by eye in
the \HST/WFPC2 data and in Digital Sky Survey images, and it also
appears in subsequently released Sloan Digital Sky Survey (SDSS) data.

\begin{figure}

  \centerline{
    \includegraphics[height=1.6in,angle=0]{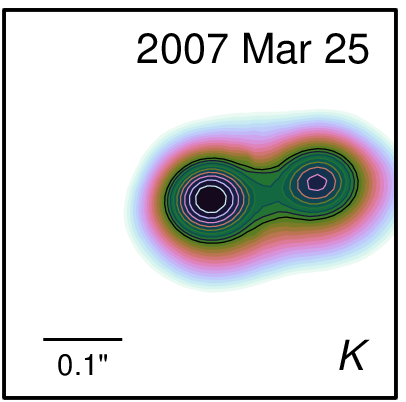} \hskip 0.04in
    \includegraphics[height=1.6in,angle=0]{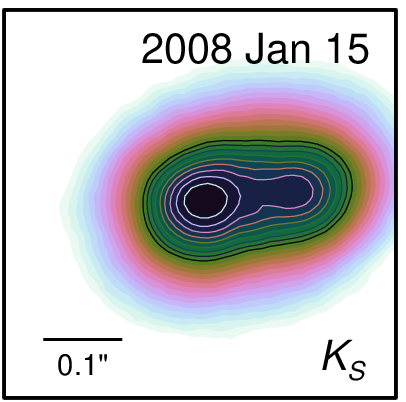} \hskip 0.04in
    \includegraphics[height=1.6in,angle=0]{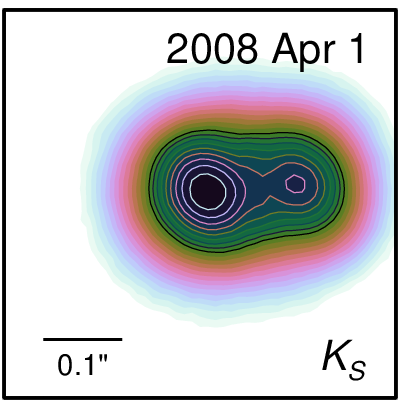} }
  \vskip 0.05in
  \centerline{
    \includegraphics[height=1.6in,angle=0]{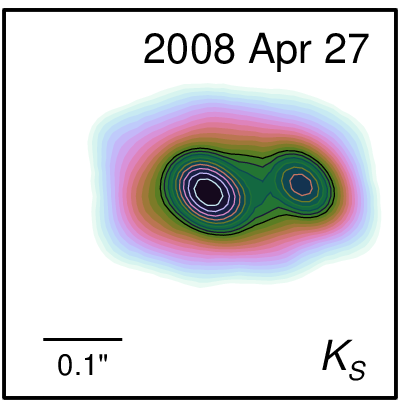} \hskip 0.04in
    \includegraphics[height=1.6in,angle=0]{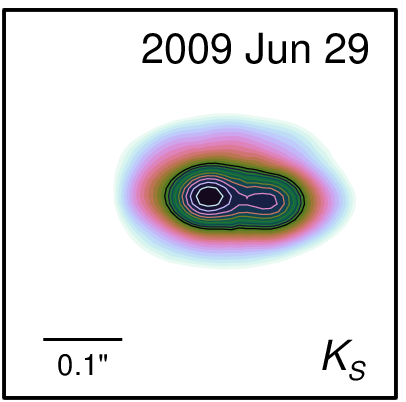} \hskip 0.04in
    \includegraphics[height=1.6in,angle=0]{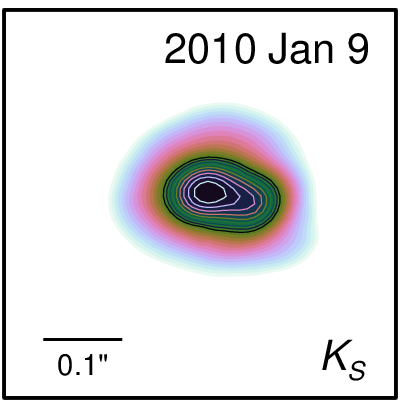} }
  \vskip 0.05in
  \centerline{
    \includegraphics[height=1.6in,angle=0]{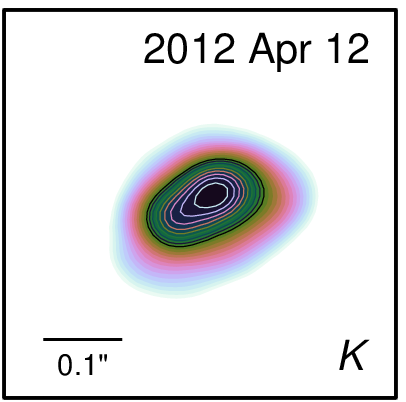} \hskip 0.14in
    \includegraphics[height=1.6in,angle=0]{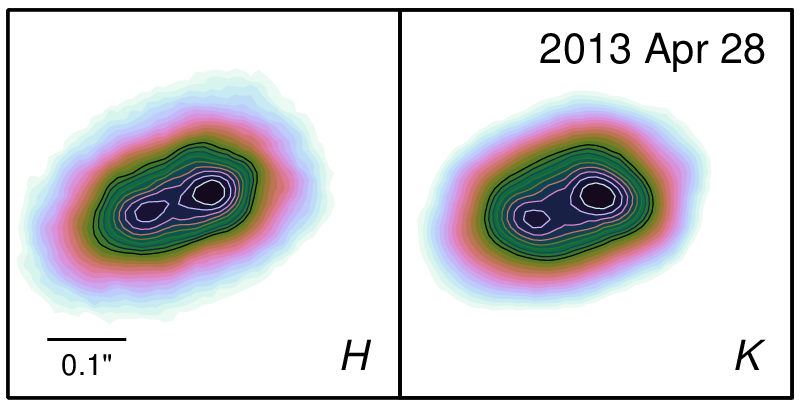} }
  \vskip 0.05in
  \centerline{
    \includegraphics[height=1.6in,angle=0]{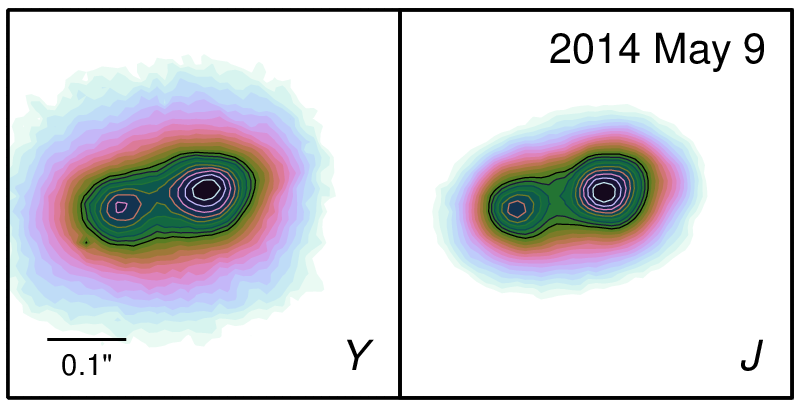} }

  \caption{\normalsize Contour plots of our Keck LGS AO images from
    which we derive astrometry and flux ratios (Table~\ref{tbl:obs}).
    Contours are in logarithmic intervals from unity to 7\% of the
    peak flux in each band.  The image cutouts are all the same size
    and have the same native pixel scale, and we have rotated them
    such that north is up for display purposes. \label{fig:keck}}

\end{figure}

Our procedure for reducing and analyzing Keck LGS data is described in
detail in our previous work \citep{2009ApJ...692..729D,
  2009ApJ...699..168D, 2010ApJ...721.1725D}. To summarize briefly, we
measure binary parameters using the StarFinder software package
\citep{2000A&AS..147..335D} when the separation is large and by
fitting three-component Gaussians when the components are too close
for StarFinder to robustly identify two distinct sources. We derive
uncertainties by applying our fitting method to artificial binary
images constructed from images of PSF reference stars with similar
FWHM and Strehl ratio, as well as by checking the scatter between
individual dithered images.  We use the NIRC2 astrometric calibration
from \citet{2010ApJ...725..331Y}, which includes a correction for the
nonlinear distortion of the camera and has a pixel scale of
$9.952\pm0.002$\,mas\,\perpix\ and an orientation for the detector's
$+y$-axis of $+0\fdg252\pm0\fdg009$ east of north.
Figure~\ref{fig:keck} shows contour plots of our imaging data at each
epoch, stacked for the purposes of display.

In Table~\ref{tbl:obs}, we present the results of our Keck imaging,
including the FWHM and Strehl ratio at each epoch along with the
derived binary parameters. As a check on these parameters, we note
that the \Ks-band flux ratio is consistent between the five epochs for
which we have data in that bandpass with a $\chi^2$ of 4.8 for 4
degrees of freedom (dof; $p = 0.31$). The weighted average and
corresponding error is $\Delta\Ks = 0.413\pm0.020$\,mag.

\subsection{\HST/WFPC2 \label{sec:hst}}

The \HST/WFPC2 discovery images of \glBC\ were taken on
2001~February~14~UT as part of GO-8581 (PI Reid), with one image each
in F814W and F1042M.  We used the F814W image for measuring
astrometry, because the PSF is smaller at shorter wavelengths and thus
enables more robust deblending of this tight 1.6-pixel binary.  As in
our previous work \citep[e.g.,][]{2008ApJ...689..436L,
  2009ApJ...706..328D}, we applied a binary fitting routine based on
the TinyTim model of the \HST\ PSF \citep{1995ASPC...77..349K} to both
the data and numerous simulated binary images constructed from single
stars observed by \HST/WFPC2.  The simulated binaries were constrained
to be within 0.2~pixels of the actual configuration of \glBC, and the
difference between the input and output values were used to determine
the rms scatter and systematic offsets in the binary parameters.  We
found the offsets to be somewhat less than the scatter in separation
($-1.9\pm2.3$\,mas), position angle (P.A.; $+0\fdg6\pm1\fdg5$), and
flux ratio ($+0.04\pm0.07$\,mag), where our quoted values are ${\rm
  offset} \pm {\rm rms}$.

Table~\ref{tbl:obs} shows the derived binary parameters after applying
the offsets from our simulations and using the rms values for the
errors. Our results are somewhat inconsistent with the parameters
reported in \citet{2003AJ....126.1526B} by 6.1\,mas (1.7$\sigma$),
3$\fdg$4 (1.6$\sigma$), and 0.52\,mag (4.0$\sigma$).  However, we note
that their paper adopted a simplification in determining errors,
namely that a single value for the uncertainty in given parameter was
used for all 60--150\,mas (1.3--3.3~pixel) binaries. This obscures the
fact that the tighter binaries over this range should have larger
uncertainties than the wider ones.  For example, their Figures~2 and 3
illustrate this well, and their larger truth-minus-fitted scatter over
the narrower 60--70\,mas separation range is sufficient to explain the
apparent discrepancies between our two sets of binary parameters.
There is much larger scatter in P.A.\ and flux ratio than is accounted
for by their single uncertainty values for these parameters, and there
is an apparent trend for these tightest binaries that their fitting
overestimates the separation by $\approx$2--3\,mas. In the following,
we will conservatively consider both our own \HST/WFPC2 astrometry
that relies on simulations tailored specifically to \glBC\ as well as
the values reported by \citet{2003AJ....126.1526B}.

%----------------------------------------------------------------------%

\section{Orbital Parameters of \glBC \label{sec:orbit}}

The relative astrometry for \glBC\ presented in Table~\ref{tbl:obs}
spans more than 13~years in time and 334$\degree$ in P.A., enabling us
to robustly determine its orbital parameters for the first time. As in
our previous work \citep[e.g.,][]{2008ApJ...689..436L,
  2009ApJ...692..729D}, we use a Markov Chain Monte Carlo (MCMC)
technique to determine the posterior distributions of all quantities.
We briefly summarize our procedure, as we have made some minor
modifications to our code compared to our previous work. As before, we
use a Metropolis-Hastings jump acceptance criterion with Gibbs
sampling that alters only one parameter at each step in the chain
\citep[e.g., see][]{2005AJ....129.1706F}. Before running our science
chains, we first run a test chain according to the method outlined by
\citet{2006ApJ...642..505F} in order to determine optimal step sizes
for each parameter (see Section~2.4 of \citealp{2012ApJS..201...19D}
for more details). We then run 20 science chains starting at different
points in parameter space, chosen by adding Gaussian noise scaled by
the step sizes to the best-fit values. We find the best-fit values in
advance by using our least-squares minimization routine based on the
MPFIT IDL package \citep{2009ASPC..411..251M}, as described in
\citet{2010ApJ...721.1725D}.

Each of our 20 Markov chains has $10^7$ steps, with every hundredth
step saved. We chose to step in parameters that would result in
appropriate uninformative (i.e., flat) prior assumptions.  For the
orbital period ($P$) and semimajor axis ($a$), our prior is log-flat
as we stepped in $\log{P}$ and $\log{a}$.  The priors in eccentricity
($e$), argument of periastron ($\omega$), time of periastron passage
($T_0$), and P.A.\ of the ascending node ($\Omega$) are linear-flat as
we stepped in $\sqrt{e}\cos{\omega}$, $\sqrt{e}\sin{\omega}$, $T_0$,
and $\Omega$.  Finally, our inclination ($i$) prior assumes randomly
distributed viewing configurations by stepping in $\cos{i}$.  In our
previous work we modified the parameters being stepped along after
every $5\times10^5$ iterations in order to increase the efficiency of
exploring of a ``curved'' region of low $\chi^2$ in parameter space.
However, for well-determined orbits like the case of \glBC\ this is
not necessary and the increased computational time needed for such
optimization ultimately results in lowered efficiency.

\begin{figure}
  \centerline{
    \hskip -0.2in
    \includegraphics[width=3.8in,angle=0]{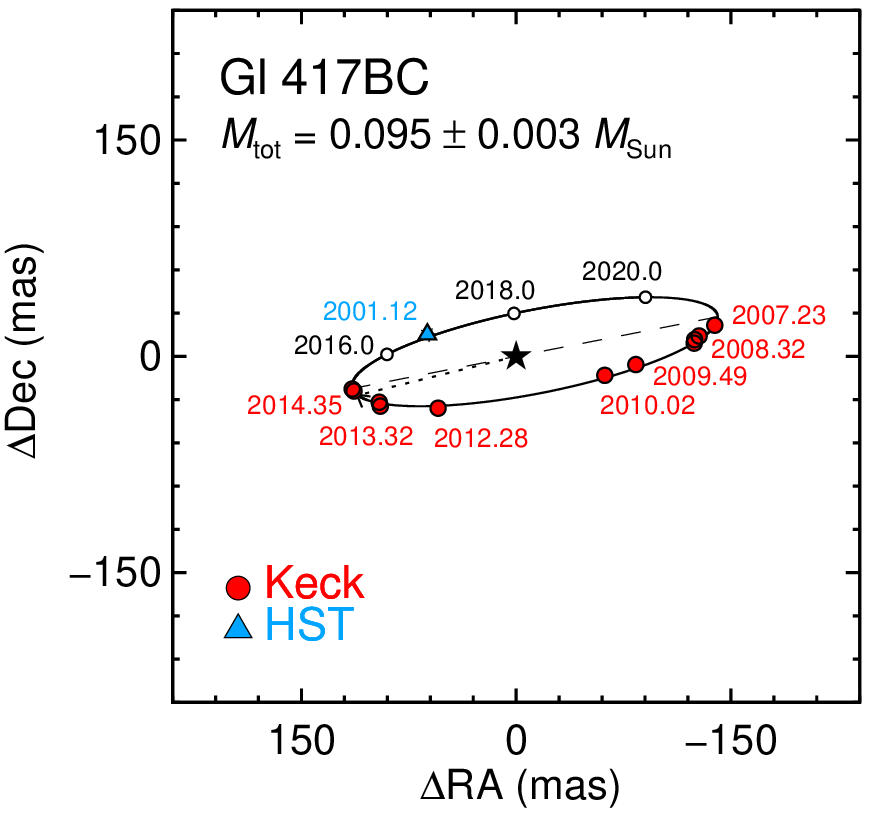}
    \hskip  0.2in
    \includegraphics[width=2.4in,angle=0]{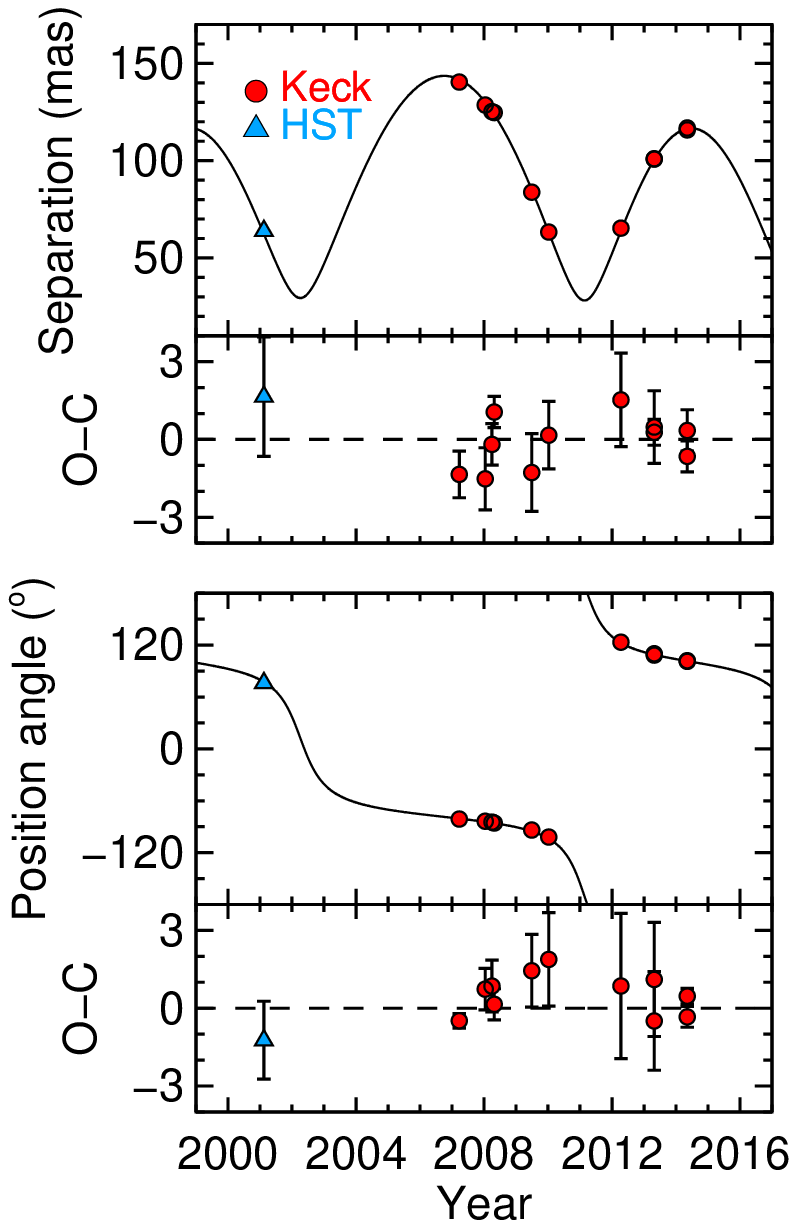} }

  \caption{\normalsize \emph{Left:} Relative astrometry for \glBC\
    along with our best-fit orbit. Error bars for the data are smaller
    than the plotting symbols. The short dotted line indicates the
    time of periastron passage, the long dashed line shows the line of
    nodes, and small empty circles show predicted future locations.
    \emph{Right:} Measurements of the projected separation and P.A.\
    of \glBC.  The best-fit orbit is shown as a solid line.  The
    bottom panels show the observed minus computed ($O-C$)
    measurements with observational error bars.\label{fig:orbit}}

\end{figure}

Table~\ref{tbl:orbit} shows the resulting orbital parameters found by
our MCMC analysis, and Figure~\ref{fig:orbit} shows our relative
astrometry alongside the best-fit orbit.  The lowest $\chi^2$ value
in our chains is 21.56 (17~dof), which has a probability of 0.202
according to the $\chi^2$ distribution, implying that our adopted
astrometric errors are reasonable.  The parameter values at this chain
step are considered the best-fit orbit, which is identical to that
found by our MPFIT routine.  In Table~\ref{tbl:orbit}, we give these
best-fit values along with the median and 1$\sigma$ and 2$\sigma$
confidence intervals for each parameter chain.  We define our
confidence intervals as the smallest range of values that captures
68.3\% and 95.4\% of the distribution.  We adopt this approach in
order to report the most likely range of parameter values from
distributions that are often asymmetrical, sometimes sharply bounded
(e.g., $0 \leq e < 1$), and in principle could be multi-modal.  In
such cases, intervals computed from the 68.3\% and 95.4\% of chain
steps centered on the median (like we have reported in previous work)
may not capture the peak of the posterior distribution.  We note that
the confidence intervals we report here all include the best-fit
parameter values.  We used the ensemble of chains to compute
Gelman-Rubin statistics for all parameters and check for convergence
and found Gelman-Rubin values of $\lesssim$1.001.  These imply that
all parameters are converged given the standard criterion requiring
Gelman-Rubin values $<$1.2 \citep{2005AJ....129.1706F}.

Combining our orbit with the parallax distance allows us to measure
the total mass (\Mtot) of \glBC\ directly from Kepler's third law. We
find $\Mtot = 0.095\pm0.003$\,\Msun\ (3.2\% error), with 1.3\% of the
error coming from the uncertainty in the orbital parameters and 2.9\%
from the parallax uncertainty. We investigated the impact of using the
different \HST/WFPC2 astrometry discussed in Section~\ref{sec:hst} by
computing additional MCMC chains. When we used the separation and
P.A.\ as reported by \citet{2003AJ....126.1526B} for the discovery
epoch, all resulting orbital parameter distributions agreed well, with
the 1$\sigma$ confidence intervals overlapping in all cases and a
minimum $\chi^2$ value nearly identical to ours. For a direct
comparison of the total mass implied by the different sets of
astrometry, we ignore the parallax error and find that the 1$\sigma$
mass interval using the \citet{2003AJ....126.1526B} astrometry
overlaps with our 1$\sigma$ range of 0.0934--0.0959\,\Msun. Thus, the
choice of \HST/WFPC2 astrometry does not have a significant impact on
the resulting dynamical mass, likely because the orbit fit is
dominated by our more precise and more numerous Keck LGS AO
astrometric data.

%----------------------------------------------------------------------%

\section{The Age and Composition of the Gl~417 System \label{sec:age}}

The age and composition of all three components in the Gl~417 system
can be established from the solar-type primary star \glA\ under the
conservative assumption that the system formed coevally from the same
bulk material. \citet{2005ApJS..159..141V} report a slightly
super-solar metallicity of $[$Fe/H$] = 0.09\pm0.03$\,dex. In the
following analysis, due to a lack of substellar evolutionary and
atmospheric models that sample metallicity at such a fine level, we
will compare to solar metallicity models but also consider the impact
of this limitation on our resulting interpretations.

Numerous methods for estimating the age of a solar-type star such as
\glA\ are available. Foremost among these in terms of claimed
precision is gyrochronology, which relies on the fact that stars lose
angular momentum with age in a predictable way
\citep[e.g.,][]{1972ApJ...171..565S}.  Recent advances in obtaining
rotation periods and membership data for large samples of stars in
clusters has enabled a calibration of this spin-down against cluster
isochronal ages, one of the most trusted clocks in astrophysics.
\citet{2007ApJ...669.1167B} defined an empirical formalism to account
for the fact that the rate of change in stellar rotation period
depends on mass, which he parametrized as a function of $(B-V)_0$
color. We have used these relations with the improved calibration from
\citet{mam08-ages} to derive the age of \glA. For the rotation period,
we use the five independent measurements obtained by
\citet{2000AJ....120.1006G} over six years. The weighted average and
rms of these rotation periods is $8.27\pm0.17$~days, which agrees well
with the period of 8~days reported by
\citet{1996ApJ...457L..99B}. Combined with the $B-V$ color of \glA\
\citep[$0.600\pm0.010$\,mag, neglecting
reddening;][]{1994cmud.book.....M}, the gyrochronology relation yields
an age of $\logtyr = 8.87\pm0.08$\,dex ($750^{+150}_{-120}$\,Myr),
where we have computed errors in a Monte Carlo fashion as described in
Section~4.3 of \citet{2009ApJ...692..729D}.

Perhaps the next best calibrated empirical relations for determining
age are those that track magnetic activity either using \ion{Ca}{2} H
and K chromospheric emission or X-ray emission. Using data from Mount
Wilson that spans approximately twenty years,
\citet{1996ApJ...457L..99B} report a time-averaged value of $\logRHK =
-4.422$\,dex for \glA. This value is in good agreement with the
single-epoch value of $\logRHK = -4.368$\,dex reported by
\citet{2003AJ....126.2048G}, as well as the older Mount Wilson value
of $\logRHK = -4.40$\,dex from \citet{1985AJ.....90.2103S}. We use the
\citet{1996ApJ...457L..99B} \logRHK\ value to compute an age for \glA\
from the chromospheric activity relations derived by
\citet{mam08-ages}. We use the method recommended by these authors
which first converts the \logRHK\ value to a Rossby number (i.e.,
rotation period divided by the convective turnover timescale \tauc),
which we find to be $0.52\pm0.10$, adopting an uncertainty of 0.10 as
suggested by \citet{mam08-ages} when using high quality Mount Wilson
data. The Rossby number is then converted to a rotation period by
estimating \tauc\ from the star's color, and this step yields $\tauc =
9.1$~days and thus an activity-derived rotation period of
$4.7\pm0.9$~days. This is finally converted to an age via their
gyrochronology relation, which gives $\logtyr = 8.44\pm0.18$\,dex.
This is strikingly different from the age derived directly from the
actual rotation period (2.2$\sigma$, given the adopted
uncertainties). We note that if we use Equation~3 from
\citet{mam08-ages}, which gives a simpler relation just between
\logRHK\ and age, we find a consistent but less precise age ($\logtyr
= 8.55\pm0.25$\,dex) that is still 1.2$\sigma$ younger than the
gyrochronology age.

\citet{mam08-ages} provide a similar method for estimating age from
X-ray emission. For \glA, \citet{1995A&A...294..515H} found an X-ray
to bolometric luminosity ratio of $\logRX \equiv \log(L_{\rm X}/\Lbol)
= -4.60\pm0.06$\,dex. Using the method proposed by \citet{mam08-ages},
we derive a Rossby number of $0.68\pm0.25$ from this \logRX\ value,
and thus an activity-derived rotation period of $6.2\pm4.9$~days. This
agrees well with the actual measured rotation period and thus results
in an age of $\logtyr = 8.72\pm0.67$\,dex that is consistent with the
gyrochronology age.

Other indicators provide some information about the age of \glA,
though not at the same precision as rotation and activity.
\citet{1981ApJ...248..651D} measured a lithium abundance of
$\log{N_{\rm Li}} = 2.38$\,dex ($W_{\rm Li} = 0.081$~\AA) for \glA.
Thus, given its effective temperature \citep[$\Teff =
5898$\,K;][]{2005ApJS..159..141V}, \glA\ lies somewhat below (i.e.,
older than) the mean relations of $\log{N_{\rm Li}}$ versus \Teff\ for
members of the 625-Myr-old Praesepe and Hyades clusters and
significantly below the 125-Myr-old Pleiades cluster
\citep{1993AJ....106.1080S, 1993AJ....105.2299S}. A comparison of
\glA's fundamental properties to stellar evolution models could offer
an isochronal age, but as is the case for many field stars the
analysis of \citet{2007ApJS..168..297T} gives only an upper limit for
the age of \glA\ ($<$2.9~Gyr, 68.3\% confidence). Finally, we compute
the heliocentric space motion of \glA, $(U, V, W) = (-15.99\pm0.17,
-23.31\pm0.23, -11.40\pm0.12)$\,\kms, with $U$ positive toward the
galactic center, based on the new \Hipparcos\ parallax and proper
motion from \citet{2007A&A...474..653V} and a radial velocity of
$-3.68\pm0.10$\,\kms\ compiled by \citet{2012A&A...546A..61D}.  The
space motion of \glA\ has not linked it to any known moving groups or
associations \citep[e.g.,][]{2000AJ....120.1006G}, and we also find no
such linkages using the online calculators of
\citet{2013ApJ...762...88M} and \citet{2014ApJ...783..121G}.

We briefly note that \glBC\ itself could provide a system age
constraint from spectroscopic signatures of low surface gravity.
\citet{2008ApJ...689.1295K} discussed this system in detail, and
without resolved optical spectroscopy they made the reasonable
assumption that the integrated-light spectrum was dominated by \glB\
($\Delta$F814W = $0.55\pm0.07$\,mag).  They found that \glB\ is
practically indistinguishable from normal, older field L4--L5 dwarfs.
The alkali lines that typically weaken for low gravity objects appear
normal, with the possible exception of \ion{Rb}{1} (7800\,\AA,
7948\,\AA).  They also note the TiO bandhead at 8200\,\AA\ might be
weaker than normal, though we note this could also be due to dilution
from the later type secondary component.  \citet{2008ApJ...689.1295K}
conclude that \glB\ is only ``slightly peculiar'' because the
spectroscopic signatures are not as obvious as in lower surface
gravity objects.  \citet{2013ApJ...772...79A} examined the
near-infrared integrated-light spectrum of \glBC\ and assigned a field
gravity classification (\textsc{fld-g}) because multiple
gravity-sensitive features were consistent with normal field objects.
Overall, we find that the spectrum of \glBC\ agrees with our older
gyrochronology age of $750^{+150}_{-120}$\,Myr and is somewhat
inconsistent with the originally published age estimate of
80--300\,Myr from \citet{2001AJ....121.3235K}.

In summary, the most robust age available for the Gl~417 system is
from gyrochronology of the solar-type primary, $\logtyr =
8.87\pm0.08$\,dex, which implies that \glA\ is somewhat older or
consistent with the Hyades.  Only one age dating method is apparently
inconsistent with this. The chromospheric emission of \glA\ traced by
\ion{Ca}{2} H and K seems to imply a much younger age of $\logtyr =
8.32\pm0.21$\,dex according to the calibration of \citet{mam08-ages}.
However, examining the range of activity levels for members of young
clusters reveals that \glA, with $\logRHK = -4.422$\,dex, is actually
comfortably within the 68\% confidence intervals of both the
500-Myr-old Ursa Majoris group ($\logRHK = -4.39$\,dex to
$-4.57$\,dex) and the 625-Myr-old Hyades cluster ($\logRHK =
-4.38$\,dex to $-4.56$\,dex). Thus, the chromospheric activity traced
by \ion{Ca}{2} H and K is not actually inconsistent with all the other
indicators, including activity traced by X-ray emission, that agree
with the gyrochronology age of $750^{+150}_{-120}$\,Myr. This agrees
with chromospheric and X-ray activity being manifestations of a
magnetic dynamo that is driven by rotation, which is also likely why
rotation--age relations show less scatter than activity--age
relations.

%----------------------------------------------------------------------%

\section{Spectral Types and Bolometric Luminosities \label{sec:props}}

\citet{2012ApJS..201...19D} report resolved spectral types for the
components of \glBC\ from spectral decomposition using its
integrated-light infrared spectrum along with the measured $K$-band
flux ratio (the March 2007 data in our Table~\ref{tbl:obs} are the
same as in their Table~5). \citet{2012ApJS..201...19D} found infrared
types of L$4.5\pm1$ and L$6\pm1$ for the primary and secondary
components, respectively, in good agreement with the integrated-light
optical spectral type of L4.5 reported by \citet{2000AJ....120..447K}.
Our new data presented here would not significantly better constrain
the resolved spectral types and are in good agreement with the
\citet{2012ApJS..201...19D} estimates of flux ratios in other
bandpasses ($\Delta{J} = 0.26\pm0.34$\,mag, $\Delta{H} =
0.32\pm0.15$\,mag), so we simply adopt the \citet{2012ApJS..201...19D}
spectral types here.

To derive bolometric luminosities (\Lbol) for both components, we used
our resolved $K$-band photometry and the bolometric
correction--spectral type relation from \citet{2010ApJ...722..311L}.
We converted the integrated-light photometry of \glBC\ from 2MASS
\citep{2003tmc..book.....C} to the MKO system using synthetic
photometry derived from the SpeX prism spectrum of \glBC\ itself
\citep{2010ApJ...710.1142B}. The resulting $K$-band photometry is
$13.29\pm0.03$\,mag and $13.63\pm0.03$\,mag for \glB\ and \glC,
respectively. When calculating bolometric corrections we account for
both the uncertainties in the resolved spectral types and scatter in
the polynomial relations of \citet{2010ApJ...722..311L}, the latter of
which dominates, and we find BC$_K = 3.31\pm0.08$\,mag for \glB\ and
BC$_K = 3.27\pm0.09$\,mag for \glC. Therefore, we arrive at
$\log(\Lbol/\Lsun)$ values of $-4.06\pm0.04$\,dex and
$-4.18\pm0.04$\,dex for the two components, respectively, where the
uncertainty in the distance is negligible compared to the
uncertainties in bolometric corrections. This corresponds to an
integrated-light bolometric flux of $\Lbol = (1.54\pm0.10) \times
10^{-4}$\,\Lsun. As a check, we computed the flux over the wavelength
range of 0.80--2.55\,\micron\ from the integrated-light SpeX spectrum
of the binary and found $L_{\rm NIR} = (0.84\pm0.02) \times
10^{-4}$\,\Lsun, which is $55\%\pm4\%$ of our total estimated
\Lbol. In comparison, for a BT-Settl model \citep{2011ASPC..448...91A}
with properties similar to the components of \glBC\ that we derive in
Section~\ref{sec:model}, $\Teff = 1700$\,K and $\logg = 5.0$\,dex
(cgs), 54\% of the bolometric flux emerges over
0.80--2.55\,\micron. Thus, our estimated \Lbol\ values are in good
agreement with direct integration of the near-IR spectral energy
distribution.

A summary of all the measured quantities for \glBC\ quoted above are
summarized in Table~\ref{tbl:meas}.  In the following analysis, we
track the covariance in luminosity ratio with other parameters like
mass ratio and the temperature difference between the two binary
components, all of which depend commonly on the uncertainties in
distance and bolometric correction.  For consistency, we also
recalculate the bolometric luminosities of the components of \hdBC\ in
the same fashion as \glBC\ described above. Using the photometry and
spectral types from \citet{2009ApJ...692..729D} we find
$\log(\Lbol/\Lsun)$ values of $-3.81\pm0.03$\,dex and
$-3.89\pm0.03$\,dex for \hdB\ and \hdC, respectively, only 0.01\,dex
different from our previously published values but with smaller errors
thanks to the improved bolometric correction relation from
\citet{2010ApJ...722..311L}.

%----------------------------------------------------------------------%

\section{Model-derived Properties for \glBC \label{sec:model}}

Substellar evolutionary models predict how the properties of brown
dwarfs depend on age for a given mass and composition. Thus with a
directly measured total mass for \glBC\ and an age and composition
inferred from the primary star \glA, we can derive model-predicted
values for \Lbol, \Teff, etc. Conversely, we can use a directly
measured property like \Lbol\ along with the system mass to infer the
age from evolutionary models, or use \Lbol\ and age to infer mass.
These mirrored scenarios correspond respectively to ``mass
benchmarks'' and ``age benchmarks,'' objects for which at least two of
three fundamental properties are measured, as discussed in detail by
\citet{2008ApJ...689..436L}. In fact, since mass and luminosity can
typically both be measured to high precision (3\%--10\%) they are the
preferred pair of parameters with which to infer other properties from
models, even if age (typical precision $\gtrsim$25\%) and mass are
both available.

We consider multiple substellar evolutionary calculations in our
analysis. The Lyon Dusty models \citep{2000ApJ...542..464C} are among
the most commonly used, and they should be appropriate for the
components of \glBC\ because their mid-L spectral types imply cloud
opacity above the photosphere.  \citet[][hereinafter
SM08]{2008ApJ...689.1327S} were the first to compute evolutionary
models in which cloud opacity changes with time.  Their hybrid models
assume the photosphere smoothly transitions from cloudy to cloudless
as objects cool from effective temperatures of 1400\,K to 1200\,K. The
components of \glBC\ turn out to both be warmer than 1400\,K according
to these models, so in our case the hybrid isochrones are essentially
equivalent to the cloudy ($f_{\rm sed} = 2$) isochrones from SM08.  We
present parameters derived from the SM08 hybrid models as well as the
fully cloudy and cloud free cases.  We also consider the Lyon Cond
models \citep{2003A&A...402..701B} that assume any clouds are
completely below the photosphere. These, along with SM08 cloud free
models, should not be appropriate for \glBC\ because the lack of dust
opacity results in spectral energy distributions highly inconsistent
with L~dwarfs, but they provide a useful counterpoint to the other
extreme assumption made by Lyon Dusty and SM08 cloudy models about the
surface boundary conditions.  Finally, we include evolutionary models
from \citet{1997ApJ...491..856B} as they are still commonly used in
the literature.  These models are cloud free, use ``gray'' atmospheres
over the temperature range considered here, and also do not benefit
from updates to opacities made over the last decade.

Our method for employing the system mass and individual luminosities
of a binary to derive all other properties from evolutionary models is
described in detail in our previous work \citep{2008ApJ...689..436L,
  2009ApJ...692..729D}. Briefly, at every given age from 10\,Myr to
10\,Gyr we calculate the model-predicted component masses, as well as
\Teff, \logg, radius ($R$), lithium abundance, and near-infrared
colors, from their measured luminosities. This is done in a Monte
Carlo fashion such that we use 10$^3$ values for a component's \Lbol,
resulting in 10$^3$ mass estimates at each age. We then step through
each of the 10$^3$ \Lbol\ pairs, considering the full range of ages
for that pair, sum the component masses as a function of age, and
determine the age that matches the measured total mass by
interpolating the curve. This is also done in a Monte Carlo fashion
such that we use 10$^3$ values for the measured \Mtot\ at this
step. This results in 10$^6$ model-derived values for every parameter,
accounting for both the errors in \Lbol\ and \Mtot\ and tracking the
covariances with \Lbol\ ratio and distance appropriately.

In Table~4, we report the median, 1-$\sigma$, and 2-$\sigma$
confidence intervals of these parameter distributions, and we
summarize some key results below.
\begin{itemize}

\item {\em System age.} The \citet{1997ApJ...491..856B} give the
  youngest age for the \glBC\ system ($410\pm30$\,Myr) because
  they predict the lowest \Lbol\ at this age and mass.  The next
  youngest model-derived ages are from the SM08 hybrid and cloudy
  models ($430\pm40$\,Myr), then Lyon Dusty ($490^{+40}_{-50}$\,Myr),
  SM08 cloud free ($540^{+50}_{-40}$\,Myr), and Lyon Cond ($570\pm50$\,Myr).
  The SM08 hybrid/cloudy age is 2.5$\sigma$ younger than the
  gyrochronology age we find for \glA\ in Section~\ref{sec:age}.  Only
  the ages derived from non-gray, cloud free models are in reasonable
  agreement, at $\approx$1$\sigma$.  This is the nearly the same level
  of discrepancy that we previously observed in the \hdBC\ system,
  which coincidentally has a very similar age and mass as \glBC.  We
  discuss the implications of this finding, as well as the differences
  in the luminosity evolution predicted by these models, in more
  detail in Section~\ref{sec:problem}.

\item {\em Effective temperature and surface gravity.} The three sets
  of models we consider here give slightly different predictions for
  \Teff\ because of the different underlying model radii.  SM08 hybrid
  models have the largest radii and correspondingly predict the lowest
  temperatures and surface gravities, with \Teff\ about 40\,K cooler
  and \logg\ about 0.03--0.04\,dex lower than from Lyon Dusty models.
  According to the Dusty models, \glB\ has $\Teff = 1750\pm30$\,K and
  $\logg = 5.11^{+0.02}_{-0.03}$\,dex (cgs), whereas the secondary
  \glC\ has $\Teff = 1630^{+30}_{-40}$\,K and $\logg =
  5.07^{+0.02}_{-0.03}$\,dex (cgs).  Lyon Cond models predict values
  about 40\,K warmer and 0.04\,dex higher gravity.  The various
  model-derived temperatures are higher than
  \citet{2009ApJ...702..154S} found for five objects of similar
  spectral type (L3.5--L7) by model atmosphere fitting.  They used the
  same model atmospheres that SM08 adopt as the boundary conditions
  for their evolutionary models, but \citeauthor{2009ApJ...702..154S}
  found that most objects were best fit by 1100--1400\,K model
  atmospheres. Only two of their objects were fit well by
  1600--1800\,K model atmospheres.  This suggests systematic errors in
  either atmospheric model spectra, evolutionary model radii, or both.
  A more rigorous test would be to fit the spectra of \glB\ and \glC\
  directly, which is challenging because few spectrographs are capable
  of resolving such a tight binary.

\item {\em Mass ratio and lithium.} The model-derived mass ratios for
  \glBC\ are all near unity, as expected from the modest measured flux
  ratios.  The two most different values are from SM08 hybrid models
  ($q \equiv M_{\rm sec}/M_{\rm pri} = 0.89\pm0.04$) and the Lyon
  Dusty models ($0.93\pm0.03$), and these are in good
  agreement. Combined with our total system mass, the various
  model-derived mass ratios and 1$\sigma$ uncertainties imply primary
  masses of 50--56\,\Mjup\ and secondary masses of 45--52\,\Mjup.
  Lyon Dusty, Cond, and \citet{1997ApJ...491..856B} models include a
  prediction for the fraction of initial lithium that remains in these
  brown dwarfs, but even in the most extreme case (Cond, 2$\sigma$)
  \glB\ is predicted to have depleted only 35\% of its lithium.  This
  is consistent with the observation of lithium absorption in the
  integrated-light spectrum of \glBC\ \citep{2000AJ....120..447K}.

\item {\em Near-infrared colors.} All models except
  \citet{1997ApJ...491..856B} predict full $JHK\Lp$ colors for both
  components of \glBC, though we ignore the cloud free models here
  because they have extremely blue colors.  Our resolved photometry
  gives somewhat redder colors for \glC\ compared to \glB\ in $J-H$
  and $J-K$ but slightly bluer colors for \glC\ at $Y-J$ and $H-K$,
  although these differences between the two components are only
  marginally significant.  In comparison to model predictions, both
  components of \glBC\ are redder than the SM08 models but bluer than
  Lyon.  The level of disagreement between most of the observed and
  predicted colors is 0.2--0.3\,mag, typical of other brown dwarfs
  with dynamical mass measurements
  \citep[e.g.,][]{2009ApJ...692..729D, 2010ApJ...721.1725D}.  However,
  the Dusty models predict colors for \glC\ that are 0.4--1.0\,mag
  discrepant.

\end{itemize}

%----------------------------------------------------------------------%

\section{A Substellar Luminosity Problem? \label{sec:problem}}

Gl~417BC is only the second field brown dwarf system after \hdBC\ with
a precisely measured mass, age, and luminosity.  By coincidence, the
two systems have very similar fundamental properties, with the
exception of the projected separation of their host stars.  Thus they
provide two independent tests of substellar models at roughly the same
age, mass, and metallicity but with objects that may have had
different formation pathways and dynamical evolution.  This is
particularly important because our age determinations implicitly
assume that the rotational evolution of the host stars is typical
compared to single stars in open clusters, which are used to calibrate
the gyrochronology relations.  This assumption is not obvious as our
first results came from \hdBC, where the brown dwarfs lie at a
projected separation of only 47\,AU from their host star.  Such a
separation could be consistent with \hdBC\ forming via gravitational
instability in a long-lived, massive circumstellar disk
\citep[e.g.,][]{2003MNRAS.346L..36R, 2007MNRAS.382L..30S}, and since
disks are suspected to influence stellar rotation formation of a
massive binary brown dwarf in the disk may have caused atypical
rotational properties for \hdA.  But if some particular mechanism was
responsible for altering the rotational history of \hdA, it is
implausible to believe that it would also be at work in the Gl~417
system.  \glBC\ is separated from its host star by 1970\,AU in
projection, suggesting a very different dynamical history from the
HD~130948 system.

Remarkably, we find nearly the same results from these two independent
tests of substellar luminosity evolution.  As described in
Section~\ref{sec:model}, the model-derived ages for \glBC\ are
$\approx$2$\sigma$ younger than the gyro age for \glA.  This implies
that the components of \glBC\ are more luminous than expected given
their masses and age.  For comparison, we present newly derived
parameters for \hdBC\ in Table~5, using the updated orbit from
\citet{me-ecc} and improved bolometric corrections from
\citet{2010ApJ...722..311L}.  The SM08 hybrid model-derived system age
for \hdBC\ is $\logtyr = 8.59\pm0.03$\,dex, and for Lyon Dusty models
it is $8.65\pm0.03$\,dex.  These ages are 3.6$\sigma$ and 2.9$\sigma$
younger than the gyro age for \hdA, $\logtyr = 8.90\pm0.08$\,dex
\citep{2009ApJ...692..729D}.

In principle, this luminosity--age discrepancy could be caused by
systematic errors either in substellar evolutionary models or in
gyrochronology relations.  The challenges associated with age
determinations for stars are well documented \citep[e.g.,][and
references therein]{2013arXiv1311.7024S}.  Recent modeling of error
sources in the gyrochronology age for any arbitrary field star by
\citet{2014ApJ...780..159E} show that at our primary stars' masses
\citep[1.08--1.11\,\Msun;][]{2007ApJS..168..297T} and rotation periods
(7.8--8.3\,days) we may expect $\approx$20\% systematic errors in our
derived ages, comparable to our empirically derived error bars of
0.08\,dex.  However, the sources of this error (differential rotation
and varying initial conditions) should be essentially random, and thus
it would be very unlikely for two unrelated field stars to show the
same age discrepancy.  We therefore conclude that altered substellar
evolution model cooling rates would provide a simpler explanation for
our observed luminosity--age discrepancy, and in the following
analysis we proceed under this assumption. (We note that if the cause
is instead a systematic age offset present in stellar
age--rotation--activity relations, this would have its own problematic
implications that could range from incorrectly estimated ages for host
stars of directly imaged planets to fundamental errors in the ages of
the clusters used to calibrate the relations.)

\begin{figure}

  \centerline{\includegraphics[width=3.1in,angle=0]{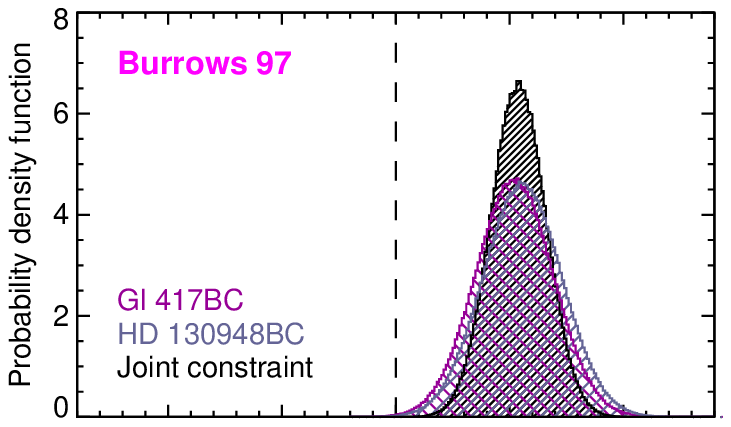}}
  \vskip -0.55in
  \centerline{\includegraphics[width=3.1in,angle=0]{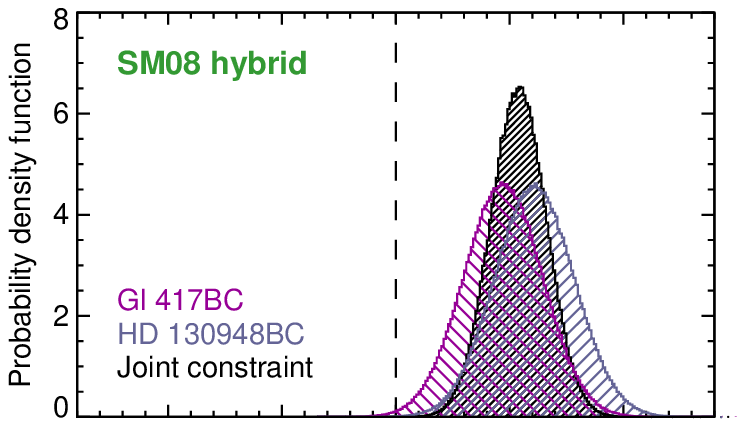}}
  \vskip -0.55in
  \centerline{\includegraphics[width=3.1in,angle=0]{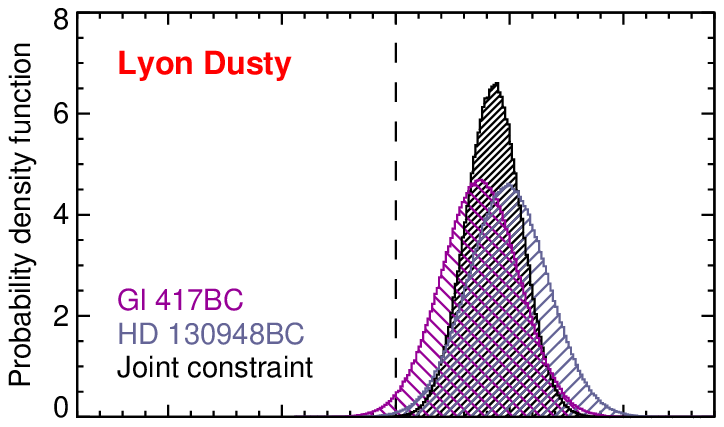}}
  \vskip -0.55in
  \centerline{\includegraphics[width=3.1in,angle=0]{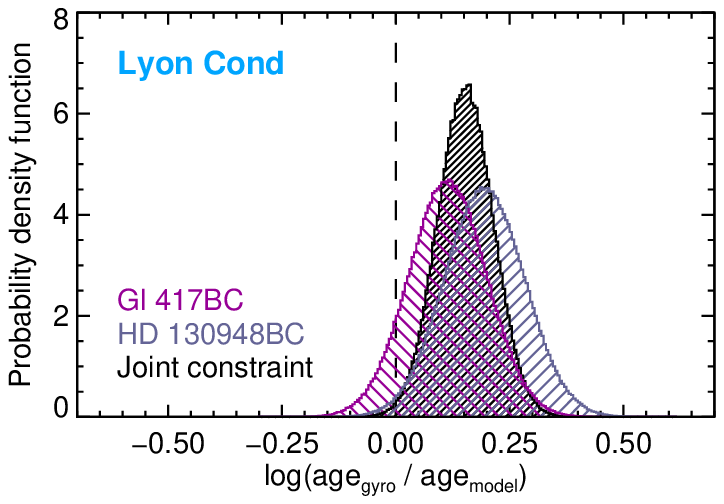}}

  \caption{\normalsize Probability distributions of the difference
    between gyrochronology ages and evolutionary model-derived ages
    for the brown dwarf binaries \glBC\ (violet) and \hdBC\ (blue).
    Multiplying these two distributions gives the joint constraint
    (black).  For both systems, all three models predict ages that are
    too young based on the measured total masses and component
    luminosities.  This indicates that model-predicted luminosities
    are too low for these binaries, which have similar component
    masses ($\approx$45--60\,\Mjup) and indistinguishable ages of
    around 800\,Myr. \label{fig:dage}}

\end{figure}

To illustrate our observed luminosity--age discrepancies, we show
probability distributions of the difference between model-derived
system ages and gyro ages in Figure~\ref{fig:dage}. We also plot the
joint probability distribution that results from combining these two
results.  The joint distribution implicitly assumes that the
components in both \hdBC\ and \glBC\ probe similar physics because
their masses are only $\approx$10\%--15\% different and their ages are
indistinguishable, $\Delta\logtyr = 0.03\pm0.11$\,dex.  The joint
discrepancy in age is 4.0$\sigma$ for SM08 hybrid models, 3.5$\sigma$
for Lyon Dusty models, and 2.3$\sigma$ for Cond.  To quantify the
discrepancy in terms of luminosity, we scaled up the \Lbol\ values
predicted by models by a constant factor to find the boost that brings
the ages into exact agreement.  For Lyon Dusty models, the scaling
factor needed was 0.25\,dex for \glBC\ and 0.35\,dex for \hdBC, and
for SM08 hybrid models they were 0.27\,dex and 0.40\,dex,
respectively.  Lyon Cond models require the smallest boost of only
0.15\,dex for \glBC\ and 0.25\,dex for \hdBC.  The fact that the level
of discrepancy varies widely between models is a reflection of the
different luminosity predictions for substellar objects in different
models.

\subsection{The Influence of Clouds on Brown Dwarf Cooling}

The inclusion of additional opacity from dust clouds has long been
known to result in a lower luminosity at a given mass and age
\citep[e.g.,][]{2000ApJ...542..464C}.  This explains why the Lyon
Dusty models are more discrepant with our unexpectedly luminous brown
dwarfs than the Cond models. SM08 also note that their cloudy models
are slightly lower luminosity than Lyon Dusty models.
Figure~\ref{fig:lbol-compare} shows the differences in predicted
luminosities as a function of age for a model object of similar mass
to the binary components we consider here (0.050\,\Msun). Across a
wide range of ages, and particularly from a few hundred Myr to
$\sim$1\,Gyr, some of the most commonly used models differ in their
luminosity predictions significantly.  At ages of 700--800\,Myr, this
amounts to a $\approx$0.2\,dex range in \Lbol.  Thus, mass estimates
from evolutionary models are actually quite dependent on the choice of
model.

A particularly interesting case shown in Figure~\ref{fig:lbol-compare}
is the one set of models that attempts to account for cloud evolution
as a substellar object cools.  The SM08 hybrid models show a feature
in \Lbol\ vs.\ age not seen in any other models: a bump in \Lbol\ that
accompanies the disappearance of cloud opacity from the photosphere.
This is at least partly understandable because an object with no cloud
opacity is more luminous, as noted above in the comparison between
Lyon Cond and Dusty models.  However, the SM08 hybrid models actually
greatly outshine the Lyon Cond models for a few Gyr after the clouds
disappear.  This is not simply a consequence of differences in the
energy transport budgets of the two models (Lyon models account for
electron conduction that dominates $\gtrsim$2\,Gyr, while SM08 do not)
because the SM08 hybrid models actually reach a similar or somewhat
lower \Lbol\ as Cond at 10\,Gyr.  Therefore, we find that cloud
evolution can have an even more profound impact on luminosity
evolution than previously thought.

The SM08 hybrid models are the first to self-consistently calculate
substellar evolution accounting for cloud disappearance, and they
adopt a simple interpolation between cloudy and cloudless model
atmospheres to do so.  However, the same group has long pointed out
the possibility of patchy clouds \citep{2001ApJ...556..872A}, and they
have also investigated the impact of having two types of clouds in
different regions of the surface on the colors and spectra of brown
dwarfs \citep{2010ApJ...723L.117M}.  Recent observations of brown
dwarf variability, particularly in the L/T transition, now provide
strong evidence for such patchy clouds
\citep[e.g.,][]{2009ApJ...701.1534A, 2012ApJ...750..105R}.  These
opacity holes are inferred to cover a significant fraction of the
surface \citep[$\sim$10\%;][]{2013ApJ...767..173H}, and this could
alter model predictions of luminosity evolution as compared to SM08's
smooth interpolation between cloudy and cloudless.  We suggest that if
cloudless regions appeared earlier, then the luminosity bump seen in
SM08 hybrid models may occur earlier, i.e., at several hundred Myr
instead of a few Gyr.  This speculative idea could provide a solution
to the over luminosity we have observed for both \hdBC\ and \glBC.

Finally, we note that metallicity is not likely to play a significant
role in modulating \Lbol.  To illustrate this, we consider SM08
cloudless models at metallicities of $-0.3$, 0.0, and +0.3\,dex at an
age of 800\,Myr.  The super-solar models predict \Lbol\ values higher
by 0.03--0.04\,dex at masses of 0.045--0.060\,\Msun, whereas the
sub-solar models predict 0.04--0.05\,dex lower \Lbol\ for the same
mass range.  Thus, at the metallicities of the Gl~417 and HD~130948
systems, ${\rm [Fe/H]} \approx 0.0$--0.1\,dex, we find from simple
interpolation that the error in our model luminosities are
$\lesssim$0.01\,dex by assuming solar metallicity, i.e., negligible
compared to our 0.03--0.04\,dex \Lbol\ measurement errors.

\subsection{Implications for Model-Derived Masses}

\begin{figure}

  \centerline{\includegraphics[width=3.1in,angle=0]{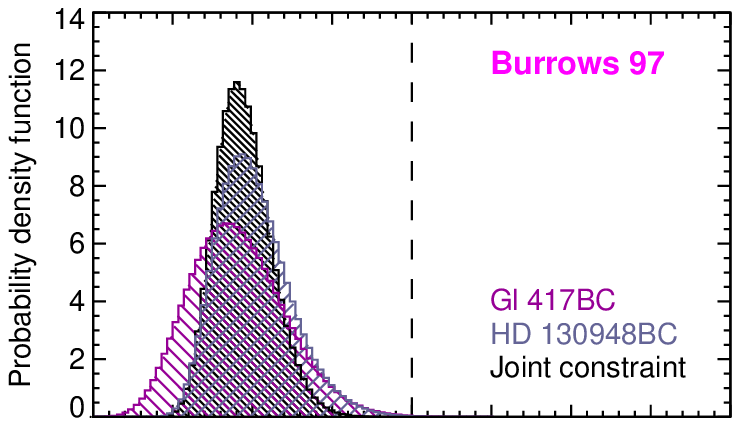}}
  \vskip -0.55in
  \centerline{\includegraphics[width=3.1in,angle=0]{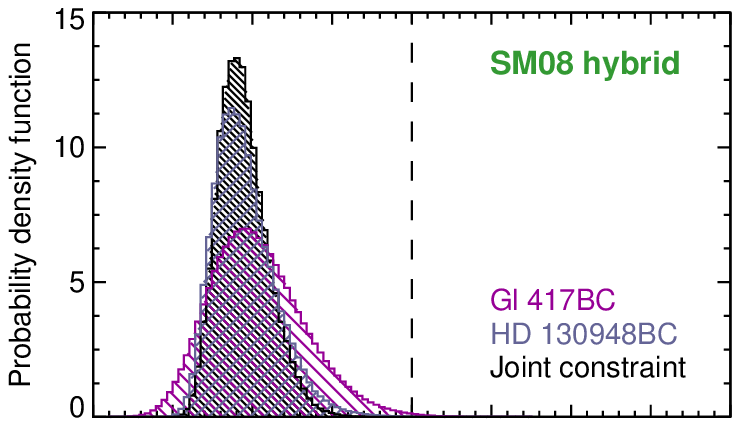}}
  \vskip -0.55in
  \centerline{\includegraphics[width=3.1in,angle=0]{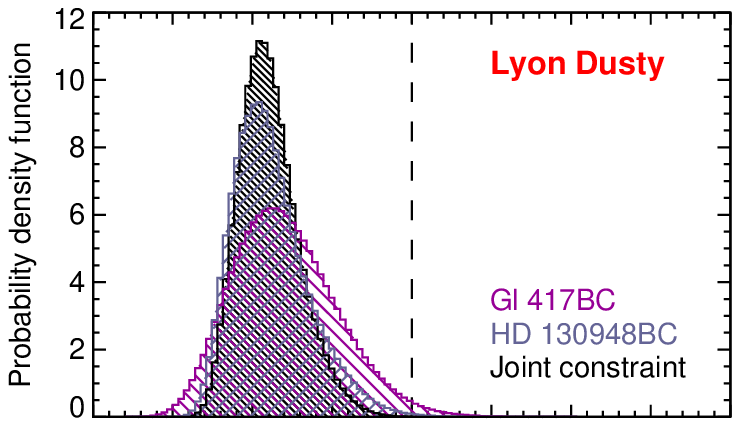}}
  \vskip -0.55in
  \centerline{\includegraphics[width=3.1in,angle=0]{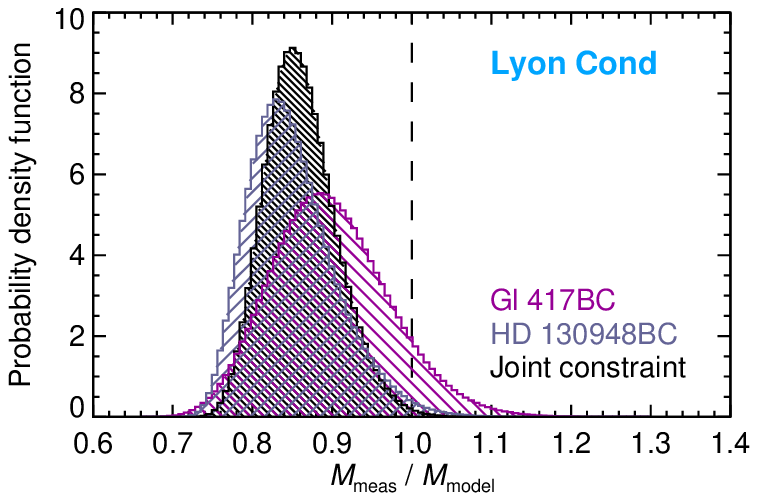}}

  \caption{\normalsize Probability distributions of the difference
    between the system masses measured dynamically and those derived
    from evolutionary models using component luminosities and system
    gyrochronology ages for the brown dwarf binaries \glBC\ (violet)
    and \hdBC\ (blue).  Multiplying these two distributions gives the
    joint constraint (black).  For both systems, the directly measured
    masses are systematically lower than predicted by all three
    models.  This is an alternative way of viewing the same
    discrepancy shown in Figure~\ref{fig:dage}, caused by
    model-predicted luminosities that are too low at this mass
    ($\approx$45--60\,\Mjup) and age
    ($\approx$800\,Myr). \label{fig:dmtot}}

\end{figure}

\begin{figure}
  \centerline{
    \includegraphics[width=3.1in,angle=0]{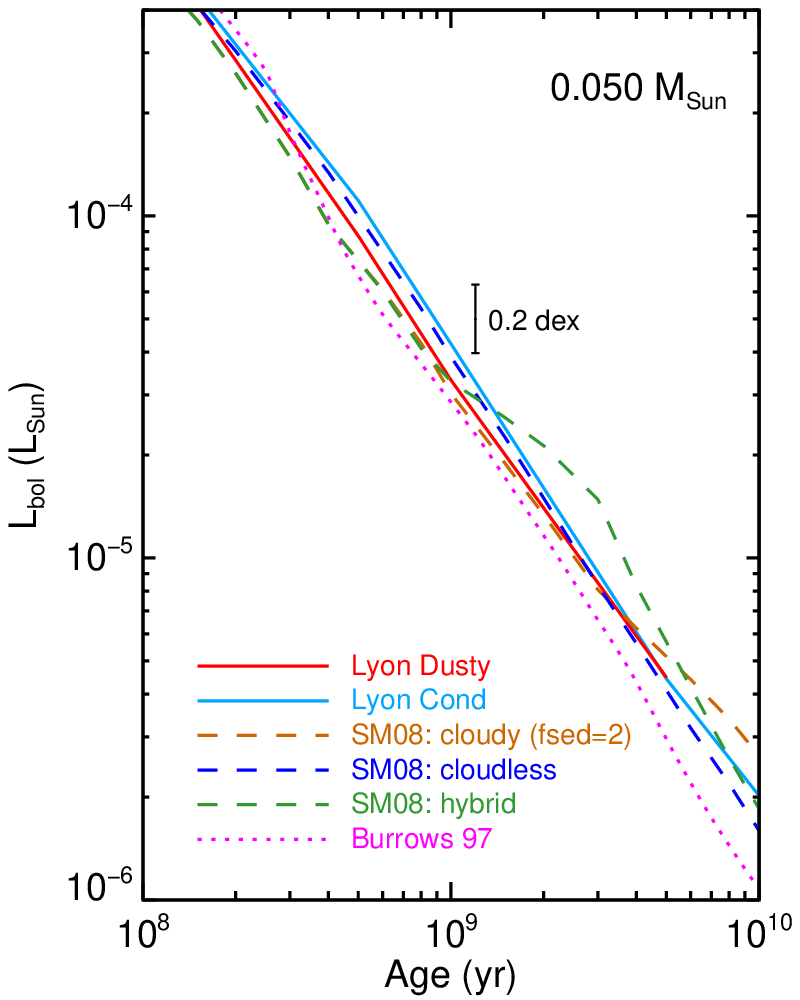}
    \hskip  0.3in
    \includegraphics[width=3.1in,angle=0]{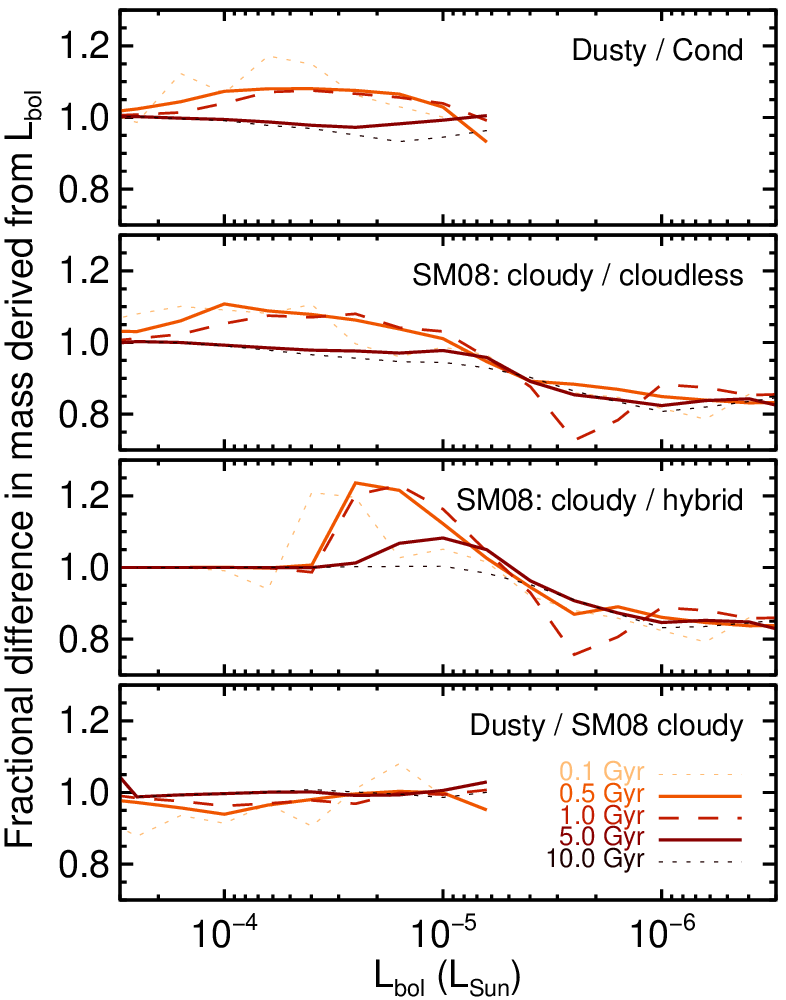}}

  \caption{\normalsize \emph{Left:} Bolometric luminosity as a
    function of age for a 0.050\,\Msun\ brown dwarf as predicted by
    several different evolutionary models: Dusty
    \citep{2000ApJ...542..464C} and Cond \citep{2003A&A...402..701B}
    models from the Lyon group; dusty ($f_{\rm sed} = 2$), cloudless,
    and hybrid models from SM08; and models from
    \citet{1997ApJ...491..856B}.  The range in model predicted
    luminosity is typically $\approx$0.2\,dex ($\approx$60\%).
    \emph{Right:} Comparison of masses that would be derived from
    evolutionary models given (errorless) \Lbol\ and age.  Each plot
    shows the ratio of the masses derived from two evolutionary
    models, e.g., the top panel shows the Dusty model-derived masses
    divided by the Cond model-derived masses for the same \Lbol\ and
    age.  The impact of clouds on luminosity evolution can result in
    model-derived masses that differ by as much as $\pm$25\%, as in
    the case of the cloudy versus SM08 hybrid
    models.  \label{fig:lbol-compare}}

\end{figure}

According to the scaling relations presented by \citet{bur01}, $\Lbol
\propto M^{2.4}$ so a large systematic error in model luminosities
will result in a correspondingly smaller error in masses derived from
those models.  To illustrate how variations in predicted luminosity
evolution impact model-derived masses, Figure~\ref{fig:dmtot} shows
the difference between our measured dynamical masses for \glBC\ and
\hdBC\ and those that we infer from evolutionary models based on the
component luminosities and system ages.  The Lyon Dusty models gives
masses 15\%--20\% ($\approx$0.09\,dex) higher than we measured, while
the discrepancy is larger for SM08 hybrid models (20\%--25\%;
$\approx$0.11\,dex).  The Lyon Cond models show the smallest \Lbol\
discrepancy, even though they are not intended to be appropriate for
these dust-bearing mid-L dwarfs, and correspondingly the masses
derived from these models are only 10\%--20\% ($\approx$0.07\,dex)
higher than we measure.

To broaden our discussion beyond \glBC\ and \hdBC, we show in
Figure~\ref{fig:lbol-compare} the fractional differences between
masses derived from the various models we consider here over a wide
range of assumed ages and \Lbol\ values.  For example, Lyon Cond
models are typically more luminous than Lyon Dusty at a given mass and
age and thus masses derived from Dusty will be higher than from Cond
at a given \Lbol\ and age.  (Note that this trend reverses at ages of
5--10\,Gyr simply because at such old ages objects have either
stabilized on the main sequence at high temperatures not significantly
affected by dust or cooled to temperatures where Dusty models are not
appropriate, $\Teff \lesssim 1000$\,K.)  Similar trends appear in the
comparison between SM08 cloudy and cloudless models that are tracked
to lower luminosities (Lyon Dusty models are not computed below $\Lbol
\approx 5\times10^{-6}$\,\Lsun).  The largest differences in
model-derived masses for the SM08 cloudy/cloudless comparison case are
$\approx$25\% and appear at 1\,Gyr and $\Lbol =
2\times10^{-6}$\,\Lsun\ ($\Teff \approx 700$\,K) because at these low
temperatures SM08 cloudy models are actually more luminous than
cloudless models.

As discussed above, the largest discrepancies between various model
predictions of luminosity evolution all involve the SM08 hybrid models
that, unlike other models, display a prominent luminosity increase as
clouds disappear.  Figure~\ref{fig:lbol-compare} shows that masses
derived from SM08 fully cloudy models, which are nearly identical to
Lyon Dusty models, are up to 25\% higher than those that would be
inferred from the SM08 hybrid models.  The SM08 cloudy/hybrid mass
discrepancy is $>$10\% over a range in \Lbol\ that corresponds to
$\Teff \approx 1000$--1400\,K and for ages up to $\approx$5\,Gyr.
Because cloud disappearance is parametrized purely by \Teff\ in the
SM08 hybrid models, the same effect occurs for younger objects but at
higher luminosities due to their larger radii.  Therefore if cloud
disappearance indeed causes such substantial changes in \Lbol\
evolution, it will affect mass estimates for substellar objects of all
ages, except perhaps at old ages of $\sim$10\,Gyr when all but the
very highest mass brown dwarfs have long been without their clouds.

%----------------------------------------------------------------------%

\section{Conclusions}

We have presented a dynamical mass measurement for the L4.5+L6 binary
\glBC\ based on Keck LGS AO imaging obtained over 2007--2014.
Combined with reanalysis of the \HST\ discovery images from 2001, our
data now span over 13 years of the $15.65\pm0.09$\,yr orbit, allowing
us to determine a precise system mass of $99\pm3$\,\Mjup.  The host
star \glA\ is a young solar-type star for which we derive a
gyrochronology age of $750^{+150}_{-120}$\,Myr that agrees with other
(less precise) ages estimated from activity indicators, lithium, and
isochrones.  \glBC\ now joins \hdBC\ as only the second system of
brown dwarfs with a precisely measured mass, age, \emph{and}
luminosity.  These two systems coincidentally have similar component
masses to within 10\%--15\%, indistinguishable ages, and nearly solar
composition.  This makes \glBC\ ideal for assessing the ``luminosity
problem'' identified by our prior work on \hdBC, for which we found
that Lyon Dusty models under-predicted the component luminosities by a
factor of $\approx$2 \citep{2009ApJ...692..729D}. Moreover, the larger
projected separation (1970\,AU) between \glBC\ and its host star
compared to \hdBC\ (47\,AU) guards against a peculiar angular momentum
history impacting stellar rotation-based age estimates.

\glBC\ displays a nearly identical over-luminosity compared to models
as we previously observed for \hdBC\ ($\Delta\log\Lbol \approx
0.3$\,dex).  This new evidence strongly suggests that there is indeed
a luminosity problem, at least for substellar objects with masses
around 45--60\,\Mjup\ at an age of $\approx$800\,Myr.  In search of a
possible solution, we compared the luminosity predictions from
currently available evolutionary models and noted that cloud
disappearance can have a surprisingly large impact on luminosity
evolution.  While it has long been recognized that cloud opacity
suppresses the luminosity of a brown dwarf at a given age and mass,
recent models actually show that as clouds disappear from a dusty
brown dwarf it can, for a time, outshine even cloud-free objects of
the same mass and age.  However, this boost to the luminosity does not
occur early enough in hybrid models from \citet{2008ApJ...689.1327S}
to explain the over-luminosity we observe.  These models adopt a
smooth interpolation between cloudy and cloudless boundary conditions,
but the latest observations of brown dwarf variability suggest a
patchy process is likely more realistic.  We therefore speculate that
opacity holes may appear early enough, e.g., at mid-L spectral types
like our two binaries, to initiate a luminosity boost that would bring
evolutionary models into agreement with our observations.

If cloud evolution is responsible for the observed luminosity problem,
evolutionary models suggest that this phase would be relatively
long-lived and span ages that encompass most of the field population
of brown dwarfs, from a few hundred Myr up to a few Gyr.  Thus,
masses derived from the commonly used dusty or cloudless evolutionary
models would be over estimated by 10\%--25\%, even for some time after
clouds disappeared from view entirely.  Many of the known directly
imaged gas-giant planets are L-type or L/T transition objects, so
their model-derived properties would be particularly susceptible to
systematic errors caused by clouds.  Under our speculative
assumptions, higher mass brown dwarfs ($>$60\,\Mjup) should not be
over-luminous at $\approx$800\,Myr because they are still too hot to
be affected by clouds, nor should brown dwarfs of similar mass to
\glBC\ and \hdBC\ that are several Gyr old, long after the \Lbol\
boost has diminished.

The most direct evidence for this proposed scenario would be to
measure the continuous mass--luminosity relation into the substellar
regime of a young cluster, which would show clearly if there is indeed
a boost in \Lbol\ associated with the disappearance of clouds.
Perhaps the most promising venue for a such a study is the Pleiades,
given that is among the nearest young clusters and the discovery of
short-period binaries suitable for dynamical mass measurements will be
enabled by high-spatial resolution surveys with \textsl{JWST} and TMT
AO.  Young moving groups might also provide such a test, though they
possess fewer members.  Precise asteroseismic stellar ages of nearby
stars from the TESS mission \citep{2010AAS...21545006R} will also
revolutionize the substellar model tests possible for companions to
stars in the solar neighborhood.  Not only will TESS data improve the
accuracy of ages for stars like Gl~417 and HD~130948, they will also
enable tests at older ages where activity-age relations are poorly
calibrated.

%----------------------------------------------------------------------%

\acknowledgments

We are grateful to 
Josh Carter for discussions about MCMC  methods; 
Didier Saumon and Isabelle Baraffe for providing expanded model grids
and helpful discussions about substellar evolution;
Joel Aycock, Randy Campbell, Al Conrad, Heather Hershley, Jim Lyke,
Jason McIlroy, Eric Nielsen, Gary Punawai, Julie Riviera, Hien Tran,
Cynthia Wilburn, and the Keck Observatory staff for assistance with
the Keck LGS AO observing;
the anonymous referee for a constructive report;
and James R.\ A.\ Davenport for distributing his IDL implementation of
the cubehelix color scheme.
This work was supported by a NASA Keck PI Data Award, administered by
the NASA Exoplanet Science Institute. 
T.J.D.\ acknowledges support from Hubble Fellowship grant
HST-HF-51271.01-A awarded by the Space Telescope Science Institute,
which is operated by AURA for NASA, under contract NAS 5-26555.
M.C.L.\ acknowledges support from NSF grant AST09-09222.
Our research has employed the 2MASS data products; NASA's
Astrophysical Data System; the SIMBAD database operated at CDS,
Strasbourg, France; and the SpeX Prism Spectral Libraries, maintained
by Adam Burgasser at \url{http://www.browndwarfs.org/spexprism}.
Finally, the authors wish to recognize and acknowledge the very
significant cultural role and reverence that the summit of Mauna Kea has
always had within the indigenous Hawaiian community.  We are most
fortunate to have the opportunity to conduct observations from this
mountain.

{\it Facilities:} \facility{Keck:II (LGS AO, NIRC2)} \facility{IRTF (SpeX)}

\clearpage

\clearpage

\clearpage

\clearpage

\clearpage

\clearpage
\begin{landscape}
\begin{deluxetable}{lccccccc}
\tablecaption{Relative astrometry and photometry for \glBC \label{tbl:obs}}
\tablewidth{0pt}
\tablehead{
\colhead{Date} &
\colhead{Airmass} &
\colhead{Filter} &
\colhead{FWHM} &
\colhead{Strehl ratio} &
\colhead{$\rho$} &
\colhead{P.A.} &
\colhead{$\Delta{m}$} \\
\colhead{(UT)} &
\colhead{} &
\colhead{} &
\colhead{(mas)} &
\colhead{} &
\colhead{(mas)} &
\colhead{(\degree)} &
\colhead{(mag)} }
\startdata
 2001-02-14                  &\nodata& F814W &      \nodata      &     \nodata     & $ 63.9\pm2.3$     & $ 76.2\pm1.5$     & $ 0.55\pm0.07 $ \\
 2007-03-25\tablenotemark{a} & 1.301 &  $K$  &     $ 91\pm5$     & $ 0.15\pm0.02 $ & $140.4\pm0.9$\phn &$278.84\pm0.28$\phn& $0.347\pm0.025$ \\
 2008-01-15                  & 1.112 &  \Ks  &     $115\pm6$ \phn& $0.074\pm0.007$ & $128.6\pm1.2$\phn & $276.3\pm0.8$ \phn& $ 0.35\pm0.04 $ \\
 2008-04-01                  & 1.331 &  \Ks  &     $112\pm4$ \phn& $0.078\pm0.011$ & $125.2\pm0.8$\phn & $275.3\pm1.0$ \phn& $ 0.41\pm0.03 $ \\
 2008-04-27                  & 1.422 &  \Ks  &     $ 85\pm4$     & $ 0.12\pm0.02 $ & $124.7\pm0.6$\phn & $274.2\pm0.6$ \phn& $ 0.47\pm0.04 $ \\
 2009-06-29                  & 1.512 &  \Ks  &     $ 68\pm2$     & $ 0.21\pm0.03 $ & $ 83.8\pm1.5$     & $266.1\pm1.4$ \phn& $ 0.47\pm0.11 $ \\
 2010-01-09                  & 1.072 &  \Ks  &     $ 66\pm3$     & $ 0.23\pm0.04 $ & $ 63.3\pm1.3$     & $258.0\pm1.8$ \phn& $ 0.43\pm0.08 $ \\
 2012-04-12                  & 1.067 &  $K$  &     $ 73\pm5$     & $ 0.20\pm0.03 $ & $ 65.3\pm1.8$     & $123.4\pm2.8$ \phn& $ 0.34\pm0.14 $ \\
 2013-04-28                  & 1.157 &  $H$  &     $ 78\pm6$     & $0.051\pm0.004$ & $100.8\pm0.5$\phn & $108.4\pm1.9$ \phn& $ 0.26\pm0.09 $ \\
                             & 1.136 &  $K$  &\phn $ 87\pm10$    & $0.111\pm0.017$ & $101.0\pm1.4$\phn & $110.0\pm2.2$ \phn& $ 0.28\pm0.13 $ \\
 2014-05-09                  & 1.107 &\Ynirc &\phn $ 78\pm10$    & $0.018\pm0.004$ & $115.8\pm0.6$\phn & $102.0\pm0.3$ \phn& $ 0.40\pm0.04 $ \\
                             & 1.097 &  $J$  &     $ 69\pm8$     & $0.047\pm0.011$ & $116.8\pm0.8$\phn & $101.2\pm0.4$ \phn& $ 0.44\pm0.04 $ \\
\enddata

\tablecomments{The first epoch of data is from \HST/WFPC2-PC, and the
  other epochs are our new Keck LGS AO measurements.  For the Keck
  images, Strehl ratios and FWHM were computed using the publicly
  available routine \texttt{NIRC2STREHL}.}

\tablenotetext{a}{This epoch was first reported in
  \citet{2012ApJS..201...19D}.}

\end{deluxetable}
\clearpage
\end{landscape}
\clearpage
\begin{deluxetable}{lcccc}
\setlength{\tabcolsep}{0.050in}
\tablewidth{0pt}
\tablecaption{Derived orbital parameters for \glBC \label{tbl:orbit}}
\tablehead{
\colhead{Parameter}   &
\colhead{Best fit}    &
\colhead{Median}      &
\colhead{68.3\% c.l.} &
\colhead{95.4\% c.l.} }
\startdata
Orbital period $P$ (yr)                         & 15.65      & 15.65      &     15.56, 15.73     &     15.49, 15.81     \\
Semimajor axis $a$ (mas)                        & 130.1      & 130.0      &     129.6, 130.5     &     129.2, 130.9     \\
Eccentricity $e$                                & 0.106      & 0.105      &     0.102, 0.109     &     0.099, 0.112     \\
Inclination $i$ (\degree)                       & 102.9      & 102.9      &     102.4, 103.4     &     101.9, 103.9     \\
P.A.\ of the ascending node $\Omega$ (\degree)  & 101.03     & 101.01     &    100.79, 101.23    &    100.57, 101.46    \\
Argument of periastron $\omega$ (\degree)       & 348        & 347        &       343, 352       &       339, 356       \\
Time of periastron $T_0-2456664.5$ (JD)         & 0          & $-$10      &     $-$70, 50\phs    &    $-$130, 110\phs   \\
Total mass (\Msun): fitted                      & 0.0949     & 0.0947     &    0.0934, 0.0959    &    0.0923, 0.0972    \\
Total mass (\Msun): final                       & 0.0949     & 0.0947     &    0.0916, 0.0976    &    0.0888, 0.1008    \\
\enddata

\tablecomments{For each parameter we report the value corresponding to
  the best fit (i.e., the lowest $\chi^2$ in the MCMC chain,
  $\chi^2_{\rm min} = 21.56$) along with the median of the posterior
  distribution and the shortest intervals containing 68.3\% and 95.4\%
  of the chain steps (i.e., 1$\sigma$ and 2$\sigma$ confidence
  limits). For clarity, the time of periastron passage is reported
  relative to the best-fit value of 2456664.5~JD (i.e., 2014~January~7
  00:00~UT).  Without resolved radial velocities there is a
  180\degree\ ambiguity in both $\Omega$ and $\omega$.  The ``fitted''
  total mass represents the results from fitting the observed orbital
  motion without accounting for the parallax error.  The ``final''
  total mass includes the additional error in the mass due to the
  error in the parallax.}

\end{deluxetable}
\clearpage
\begin{deluxetable}{lccc}
\tablewidth{0pt}
\tablecaption{Measured Properties of \glBC \label{tbl:meas}}
\tablehead{
\colhead{Property}   &
\colhead{\glB}       &
\colhead{\glC}       &
\colhead{Ref.}       }
\startdata
\Mtot\ (\Mjup)             &            \multicolumn{2}{c}{      $99\pm3$\phn   }        &  1,2  \\
\logtyr                    &            \multicolumn{2}{c}{    $8.87\pm0.08$    }        &   1   \\
Semimajor axis (AU)        &            \multicolumn{2}{c}{    $2.85\pm0.03$    }        &  1,2  \\
$d$ (pc)                   &            \multicolumn{2}{c}{   $21.93\pm0.21$\phn}        &   2   \\
Spectral type              &\phs    ${\rm L4.5}\pm1.0  $&\phs    ${\rm L6.0}\pm1.0  $    &   3   \\
$Y$ (mag)                  &\phs    $    16.37\pm 0.06 $&\phs    $    16.77\pm 0.06 $    &  1,4  \\
$J$ (mag)                  &\phs    $    15.05\pm 0.04 $&\phs    $    15.49\pm 0.04 $    &  1,4  \\
$H$ (mag)                  &\phs    $    14.19\pm 0.05 $&\phs    $    14.45\pm 0.06 $    &  1,4  \\
$K$ (mag)                  &\phs    $    13.29\pm 0.03 $&\phs    $    13.63\pm 0.03 $    &  1,4  \\
$Y-J$ (mag)                &\phs\phn$     1.32\pm0.06  $&\phs\phn$     1.28\pm0.07  $    &  1,4  \\
$J-H$ (mag)                &\phs\phn$     0.86\pm0.06  $&\phs\phn$     1.04\pm0.07  $    &  1,4  \\
$H-K$ (mag)                &\phs\phn$     0.91\pm0.06  $&\phs\phn$     0.82\pm0.07  $    &  1,4  \\
$J-K$ (mag)                &\phs\phn$     1.76\pm0.05  $&\phs\phn$     1.86\pm0.05  $    &  1,4  \\
$M_Y$ (mag)                &\phs    $    14.67\pm0.07  $&\phs    $    15.07\pm0.07  $    & 1,2,4 \\
$M_J$ (mag)                &\phs    $    13.34\pm0.04  $&\phs    $    13.78\pm0.05  $    & 1,2,4 \\
$M_H$ (mag)                &\phs    $    12.49\pm0.06  $&\phs    $    12.74\pm0.06  $    & 1,2,4 \\
$M_K$ (mag)                &\phs    $    11.58\pm0.04  $&\phs    $    11.93\pm0.04  $    & 1,2,4 \\
BC$_K$ (mag)               &\phs\phn$     3.31\pm0.08  $&\phs\phn$     3.27\pm 0.09 $    &  1,5  \\
$\log$(\Lbol/\Lsun)        &\phn    $    -4.06\pm0.04  $&\phn    $    -4.18\pm 0.04 $    &   1   \\
$\Delta\log$(\Lbol)        &              \multicolumn{2}{c}{$  0.12\pm 0.05 $}          &   1   \\
\enddata

\tablecomments{All near-infrared photometry on the MKO system, with
  $Y$-band specifically on the UKIRT system assuming that our Keck
  flux ratio $\Delta\Ynirc = \Delta{Y}_{\rm MKO}$ due to the similar
  component spectral types.}

\tablerefs{(1)~This work; (2)~\citet{2007A&A...474..653V};
  (3)~\citet{2012ApJS..201...19D}; (4)~\citet{2003tmc..book.....C};
  (5)~\citet{2010ApJ...722..311L}.}

\end{deluxetable}
\clearpage


\section*{Supplementary material}

Table 4. Evolutionary model-derived properties for Gl 417BC

| Property | Lyon Dusty (Chabrier et al. 2000) | | | Lyon Cond (Baraffe et al. 2003) | | | Tucson (Burrows et al. 1997) | | |
| --- | --- | --- | --- | --- | --- | --- | --- | --- | --- |
| | Median | 68.3% c.l. | 95.4% c.l. | Median | 68.3% c.l. | 95.4% c.l. | Median | 68.3% c.l. | 95.4% c.l. |
| Age ($t$, Gyr) | 0.49 | 0.44, 0.53 | 0.41, 0.58 | 0.58 | 0.52, 0.62 | 0.48, 0.68 | 0.41 | 0.38, 0.44 | 0.34, 0.48 |
| $\log(t/\mathrm{yr})$ | 8.69 | 8.65, 8.72 | 8.62, 8.76 | 8.76 | 8.72, 8.80 | 8.69, 8.84 | 8.61 | 8.58, 8.65 | 8.54, 8.69 |
| $M_\mathrm{B}$ ($M_\mathrm{Jup}$) | 51.9 | 50.1, 53.7 | 48.3, 55.5 | 51.9 | 50.0, 53.8 | 48.2, 55.8 | 51.5 | 49.5, 53.2 | 48.0, 55.1 |
| $M_\mathrm{C}$ ($M_\mathrm{Jup}$) | 47.4 | 45.6, 49.1 | 44.1, 51.1 | 47.4 | 45.5, 49.1 | 43.9, 51.1 | 47.8 | 46.1, 49.5 | 44.5, 51.3 |
| $q \equiv M_\mathrm{C}/M_\mathrm{B}$ | 0.91 | 0.88, 0.94 | 0.85, 0.98 | 0.913 | 0.874, 0.943 | 0.845, 0.976 | 0.929 | 0.901, 0.959 | 0.872, 0.986 |
| $T_\mathrm{eff,B}$ (K) | 1750 | 1720, 1780 | 1680, 1820 | 1790 | 1760, 1820 | 1720, 1860 | 1780 | 1740, 1810 | 1710, 1850 |
| $T_\mathrm{eff,C}$ (K) | 1630 | 1590, 1660 | 1560, 1700 | 1670 | 1630, 1700 | 1600, 1740 | 1650 | 1620, 1690 | 1580, 1730 |
| $\Delta T_\mathrm{eff}$ (K) | 130 | 80, 170 | 20, 220 | 120 | 70, 170 | 20, 210 | 130 | 70, 170 | 20, 230 |
| $\log(g_\mathrm{B})$ (cgs) | 5.103 | 5.081, 5.126 | 5.058, 5.151 | 5.142 | 5.119, 5.164 | 5.096, 5.188 | 5.142 | 5.118, 5.166 | 5.095, 5.190 |
| $\log(g_\mathrm{C})$ (cgs) | 5.063 | 5.040, 5.085 | 5.019, 5.111 | 5.105 | 5.082, 5.128 | 5.059, 5.152 | 5.105 | 5.082, 5.128 | 5.058, 5.152 |
| $R_\mathrm{B}$ ($R_\mathrm{Jup}$) | 1.010 | 0.998, 1.022 | 0.983, 1.031 | 0.957 | 0.945, 0.967 | 0.938, 0.983 | 0.959 | 0.949, 0.969 | 0.939, 0.979 |
| $R_\mathrm{C}$ ($R_\mathrm{Jup}$) | 1.010 | 1.000, 1.021 | 0.987, 1.032 | 0.957 | 0.944, 0.966 | 0.938, 0.983 | 0.964 | 0.955, 0.974 | 0.945, 0.984 |
| $(\mathrm{Li/Li_0})_\mathrm{B}$ | 0.957 | 0.922, 0.978 | 0.742, 0.982 | 0.955 | 0.905, 0.977 | 0.673, 0.980 | 0.9935 | 0.991, 0.998 | 0.974, 1.000 |
| $(\mathrm{Li/Li_0})_\mathrm{C}$ | 0.975 | 0.967, 0.985 | 0.956, 0.991 | 0.974 | 0.966, 0.983 | 0.956, 0.991 | 0.9987 | 0.998, 1.000 | 0.995, 1.000 |
| $(J-H)_\mathrm{B}$ (mag) | 1.15 | 1.03, 1.25 | 0.97, 1.39 | 0.066 | 0.046, 0.083 | 0.027, 0.101 | ⋯ | ⋯ , ⋯ | ⋯ , ⋯ |
| $(J-H)_\mathrm{C}$ (mag) | 1.61 | 1.48, 1.75 | 1.33, 1.88 | −0.001 | −0.023, 0.017 | −0.039, 0.042 | ⋯ | ⋯ , ⋯ | ⋯ , ⋯ |
| $(H-K)_\mathrm{B}$ (mag) | 0.88 | 0.78, 0.95 | 0.73, 1.06 | 0.220 | 0.207, 0.232 | 0.195, 0.247 | ⋯ | ⋯ , ⋯ | ⋯ , ⋯ |
| $(H-K)_\mathrm{C}$ (mag) | 1.25 | 1.15, 1.36 | 1.01, 1.46 | 0.204 | 0.198, 0.211 | 0.190, 0.220 | ⋯ | ⋯ , ⋯ | ⋯ , ⋯ |
| $(J-K)_\mathrm{B}$ (mag) | 2.03 | 1.82, 2.21 | 1.70, 2.46 | 0.285 | 0.254, 0.313 | 0.226, 0.344 | ⋯ | ⋯ , ⋯ | ⋯ , ⋯ |
| $(J-K)_\mathrm{C}$ (mag) | 2.86 | 2.63, 3.11 | 2.35, 3.34 | 0.202 | 0.173, 0.224 | 0.154, 0.257 | ⋯ | ⋯ , ⋯ | ⋯ , ⋯ |
| $(K-L')_\mathrm{B}$ (mag) | 1.10 | 1.05, 1.14 | 1.02, 1.20 | 1.30 | 1.26, 1.34 | 1.22, 1.37 | ⋯ | ⋯ , ⋯ | ⋯ , ⋯ |
| $(K-L')_\mathrm{C}$ (mag) | 1.31 | 1.25, 1.38 | 1.17, 1.43 | 1.42 | 1.38, 1.45 | 1.34, 1.48 | ⋯ | ⋯ , ⋯ | ⋯ , ⋯ |

Note. — Each line in the table gives the median model-derived value along with the shortest intervals containing 68.3% and 95.4% of the chain steps (i.e., $1\sigma$ and $2\sigma$ confidence limits). All photometry on the MKO system.

Table 4. (Continued) Evolutionary model-derived properties for Gl 417BC

| Property | SM08 hybrid | | | SM08 cloudy ($f_{\rm sed} = 2$) | | | SM08 cloud free | | |
|---|---|---|---|---|---|---|---|---|---|
| | Median | 68.3% c.l. | 95.4% c.l. | Median | 68.3% c.l. | 95.4% c.l. | Median | 68.3% c.l. | 95.4% c.l. |
| Age ($t$, Gyr) | 0.43 | 0.39, 0.47 | 0.36, 0.52 | 0.43 | 0.39, 0.47 | 0.36, 0.52 | 0.54 | 0.50, 0.59 | 0.45, 0.64 |
| $\log(t/{\rm yr})$ | 8.64 | 8.59, 8.68 | 8.56, 8.72 | 8.64 | 8.59, 8.68 | 8.56, 8.72 | 8.73 | 8.70, 8.77 | 8.66, 8.81 |
| $M_{\rm B}$ ($M_{\rm Jup}$) | 52.6 | 50.7, 54.5 | 48.8, 56.3 | 52.6 | 50.7, 54.4 | 48.8, 56.4 | 51.9 | 50.0, 53.7 | 48.2, 55.6 |
| $M_{\rm C}$ ($M_{\rm Jup}$) | 46.6 | 44.6, 48.6 | 42.9, 50.5 | 46.7 | 44.7, 48.6 | 43.0, 50.6 | 47.3 | 45.6, 49.1 | 43.9, 50.9 |
| $q \equiv M_{\rm C}/M_{\rm B}$ | 0.89 | 0.85, 0.92 | 0.81, 0.97 | 0.89 | 0.85, 0.93 | 0.81, 0.97 | 0.91 | 0.88, 0.95 | 0.85, 0.98 |
| $T_{\rm eff,B}$ (K) | 1710 | 1670, 1740 | 1640, 1770 | 1700 | 1670, 1740 | 1640, 1770 | 1760 | 1730, 1800 | 1700, 1830 |
| $T_{\rm eff,C}$ (K) | 1580 | 1550, 1620 | 1520, 1650 | 1580 | 1550, 1620 | 1520, 1650 | 1640 | 1610, 1680 | 1580, 1710 |
| $\Delta T_{\rm eff}$ (K) | 120 | 70, 170 | 20, 220 | 120 | 80, 180 | 30, 220 | 120 | 70, 170 | 30, 210 |
| $\log(g_{\rm B})$ (cgs) | 5.077 | 5.054, 5.102 | 5.030, 5.124 | 5.078 | 5.054, 5.101 | 5.030, 5.125 | 5.129 | 5.106, 5.151 | 5.082, 5.173 |
| $\log(g_{\rm C})$ (cgs) | 5.021 | 4.994, 5.047 | 4.972, 5.071 | 5.021 | 4.995, 5.046 | 4.971, 5.073 | 5.090 | 5.068, 5.113 | 5.045, 5.135 |
| $R_{\rm B}$ ($R_{\rm Jup}$) | 1.043 | 1.031, 1.055 | 1.020, 1.065 | 1.043 | 1.031, 1.055 | 1.019, 1.065 | 0.977 | 0.967, 0.987 | 0.957, 0.998 |
| $R_{\rm C}$ ($R_{\rm Jup}$) | 1.049 | 1.037, 1.061 | 1.027, 1.071 | 1.049 | 1.038, 1.061 | 1.026, 1.071 | 0.975 | 0.965, 0.985 | 0.956, 0.995 |
| $(Y-J)_{\rm B}$ (mag) | 1.135 | 1.122, 1.151 | 1.111, 1.162 | 1.186 | 1.180, 1.197 | 1.165, 1.199 | 1.338 | 1.330, 1.346 | 1.319, 1.351 |
| $(Y-J)_{\rm C}$ (mag) | 1.185 | 1.169, 1.198 | 1.159, 1.216 | 1.161 | 1.146, 1.179 | 1.131, 1.189 | 1.354 | 1.353, 1.355 | 1.349, 1.355 |
| $(J-H)_{\rm B}$ (mag) | 0.727 | 0.693, 0.756 | 0.669, 0.796 | 0.699 | 0.661, 0.728 | 0.637, 0.767 | 0.340 | 0.325, 0.362 | 0.304, 0.375 |
| $(J-H)_{\rm C}$ (mag) | 0.87 | 0.83, 0.91 | 0.77, 0.93 | 0.84 | 0.80, 0.87 | 0.76, 0.90 | 0.266 | 0.243, 0.291 | 0.219, 0.313 |
| $(H-K)_{\rm B}$ (mag) | 0.648 | 0.613, 0.674 | 0.591, 0.714 | 0.63 | 0.60, 0.66 | 0.57, 0.69 | 0.079 | 0.054, 0.108 | 0.029, 0.131 |
| $(H-K)_{\rm C}$ (mag) | 0.78 | 0.75, 0.81 | 0.70, 0.84 | 0.729 | 0.716, 0.748 | 0.686, 0.756 | −0.016 | −0.043, 0.013 | −0.073, 0.039 |
| $(J-K)_{\rm B}$ (mag) | 1.38 | 1.31, 1.43 | 1.26, 1.51 | 1.33 | 1.26, 1.39 | 1.21, 1.46 | 0.42 | 0.38, 0.47 | 0.33, 0.51 |
| $(J-K)_{\rm C}$ (mag) | 1.65 | 1.58, 1.72 | 1.47, 1.77 | 1.56 | 1.52, 1.62 | 1.45, 1.66 | 0.25 | 0.20, 0.30 | 0.14, 0.35 |
| $(K-L')_{\rm B}$ (mag) | 0.897 | 0.869, 0.914 | 0.863, 0.945 | 0.953 | 0.932, 0.976 | 0.908, 0.994 | 1.149 | 1.130, 1.165 | 1.113, 1.183 |
| $(K-L')_{\rm C}$ (mag) | 1.000 | 0.972, 1.026 | 0.935, 1.052 | 1.016 | 1.009, 1.024 | 0.989, 1.029 | 1.216 | 1.195, 1.236 | 1.175, 1.256 |

Note. — Saumon & Marley (2008) hybrid and cloudy models give virtually identical physical parameters because the only difference is assuming $f_{\rm sed} = 2$ for the fully cloudy models and $f_{\rm sed} = 1$ for the cloudy portion of the hybrid models.

Table 5. Evolutionary model-derived properties for HD 130948BC

| Property | Lyon Dusty (Chabrier et al. 2000) | | | Lyon Cond (Baraffe et al. 2003) | | | Tucson (Burrows et al. 1997) | | |
|---|---|---|---|---|---|---|---|---|---|
| | Median | 68.3% c.l. | 95.4% c.l. | Median | 68.3% c.l. | 95.4% c.l. | Median | 68.3% c.l. | 95.4% c.l. |
| Age ($t$, Gyr) | 0.45 | 0.42, 0.48 | 0.39, 0.51 | 0.51 | 0.47, 0.54 | 0.44, 0.58 | 0.416 | 0.391, 0.440 | 0.370, 0.466 |
| $\log(t/\mathrm{yr})$ | 8.65 | 8.62, 8.68 | 8.60, 8.71 | 8.70 | 8.67, 8.73 | 8.65, 8.76 | 8.62 | 8.59, 8.64 | 8.57, 8.67 |
| $M_\mathrm{B}$ ($M_\mathrm{Jup}$) | 58.9 | 57.3, 60.3 | 55.9, 61.9 | 59.1 | 57.6, 60.7 | 56.0, 62.3 | 58.5 | 57.1, 59.8 | 55.8, 61.1 |
| $M_\mathrm{C}$ ($M_\mathrm{Jup}$) | 55.8 | 54.4, 57.2 | 53.0, 58.7 | 55.5 | 54.0, 57.1 | 52.5, 58.7 | 56.2 | 54.9, 57.5 | 53.7, 58.9 |
| $q \equiv M_\mathrm{C}/M_\mathrm{B}$ | 0.948 | 0.920, 0.979 | 0.894, 1.013 | 0.94 | 0.91, 0.97 | 0.88, 1.01 | 0.962 | 0.938, 0.981 | 0.916, 1.003 |
| $T_\mathrm{eff,B}$ (K) | 1990 | 1960, 2030 | 1920, 2060 | 2030 | 1990, 2060 | 1960, 2090 | 2050 | 2010, 2090 | 1970, 2130 |
| $T_\mathrm{eff,C}$ (K) | 1910 | 1880, 1940 | 1840, 1970 | 1940 | 1910, 1980 | 1880, 2010 | 1960 | 1930, 2000 | 1890, 2040 |
| $\Delta T_\mathrm{eff}$ (K) | 90 | 30, 130 | $-20$, 180 | 80 | 30, 130 | $-20$, 170 | 90 | 40, 140 | $-20$, 190 |
| $\log(g_\mathrm{B})$ (cgs) | 5.144 | 5.126, 5.160 | 5.111, 5.180 | 5.171 | 5.155, 5.187 | 5.141, 5.204 | 5.201 | 5.186, 5.216 | 5.171, 5.230 |
| $\log(g_\mathrm{C})$ (cgs) | 5.123 | 5.107, 5.138 | 5.094, 5.157 | 5.150 | 5.134, 5.166 | 5.119, 5.184 | 5.184 | 5.169, 5.199 | 5.154, 5.215 |
| $R_\mathrm{B}$ ($M_\mathrm{Jup}$) | 1.017 | 1.007, 1.028 | 0.996, 1.038 | 0.987 | 0.976, 0.997 | 0.967, 1.009 | 0.955 | 0.948, 0.962 | 0.941, 0.970 |
| $R_\mathrm{C}$ ($M_\mathrm{Jup}$) | 1.016 | 1.005, 1.028 | 0.992, 1.036 | 0.982 | 0.972, 0.992 | 0.963, 1.004 | 0.955 | 0.948, 0.962 | 0.941, 0.968 |
| $(\mathrm{Li/Li_0})_\mathrm{B}$ | 0.46 | 0.32, 0.67 | 0.13, 0.75 | 0.33 | 0.19, 0.47 | 0.05, 0.61 | 0.902 | 0.873, 0.936 | 0.828, 0.965 |
| $(\mathrm{Li/Li_0})_\mathrm{C}$ | 0.72 | 0.62, 0.83 | 0.47, 0.94 | 0.65 | 0.50, 0.80 | 0.40, 0.97 | 0.950 | 0.931, 0.978 | 0.899, 0.988 |
| $(J-H)_\mathrm{B}$ (mag) | 0.72 | 0.67, 0.76 | 0.64, 0.81 | 0.180 | 0.166, 0.196 | 0.148, 0.210 | $\cdots$ | $\cdots$ | $\cdots$ |
| $(J-H)_\mathrm{C}$ (mag) | 0.82 | 0.77, 0.87 | 0.74, 0.93 | 0.143 | 0.128, 0.158 | 0.112, 0.173 | $\cdots$ | $\cdots$ | $\cdots$ |
| $(H-K)_\mathrm{B}$ (mag) | 0.561 | 0.534, 0.587 | 0.508, 0.618 | 0.299 | 0.288, 0.311 | 0.274, 0.322 | $\cdots$ | $\cdots$ | $\cdots$ |
| $(H-K)_\mathrm{C}$ (mag) | 0.64 | 0.60, 0.67 | 0.58, 0.71 | 0.275 | 0.262, 0.287 | 0.249, 0.299 | $\cdots$ | $\cdots$ | $\cdots$ |
| $(J-K)_\mathrm{B}$ (mag) | 1.28 | 1.21, 1.34 | 1.15, 1.43 | 0.479 | 0.455, 0.509 | 0.425, 0.532 | $\cdots$ | $\cdots$ | $\cdots$ |
| $(J-K)_\mathrm{C}$ (mag) | 1.46 | 1.38, 1.54 | 1.31, 1.64 | 0.418 | 0.393, 0.447 | 0.360, 0.468 | $\cdots$ | $\cdots$ | $\cdots$ |
| $(K-L')_\mathrm{B}$ (mag) | 0.927 | 0.904, 0.951 | 0.879, 0.972 | 1.03 | 0.99, 1.07 | 0.95, 1.11 | $\cdots$ | $\cdots$ | $\cdots$ |
| $(K-L')_\mathrm{C}$ (mag) | 0.977 | 0.958, 0.997 | 0.938, 1.016 | 1.12 | 1.08, 1.16 | 1.05, 1.20 | $\cdots$ | $\cdots$ | $\cdots$ |

Note. — Each line in the table gives the median model-derived value along with the shortest intervals containing 68.3% and 95.4% of the chain steps (i.e., $1\sigma$ and $2\sigma$ confidence limits). All photometry on the MKO system.

Table 5. (Continued) Evolutionary model-derived properties for HD 130948BC

| Property | SM08 hybrid | | | SM08 cloudy ($f_{sed} = 2$) | | | SM08 cloud free | | |
|---|---|---|---|---|---|---|---|---|---|
| | Median | 68.3% c.l. | 95.4% c.l. | Median | 68.3% c.l. | 95.4% c.l. | Median | 68.3% c.l. | 95.4% c.l. |
| Age ($t$, Gyr) | 0.393 | 0.366, 0.419 | 0.343, 0.446 | 0.393 | 0.366, 0.418 | 0.343, 0.446 | 0.47 | 0.44, 0.51 | 0.41, 0.54 |
| $\log(t/\mathrm{yr})$ | 8.59 | 8.57, 8.62 | 8.54, 8.65 | 8.59 | 8.57, 8.62 | 8.54, 8.65 | 8.68 | 8.65, 8.71 | 8.62, 8.74 |
| $M_{\mathrm{B}}$ ($M_{\mathrm{Jup}}$) | 58.8 | 57.4, 60.2 | 56.0, 61.7 | 58.8 | 57.4, 60.2 | 56.0, 61.7 | 59.2 | 57.6, 60.7 | 56.1, 62.3 |
| $M_{\mathrm{C}}$ ($M_{\mathrm{Jup}}$) | 55.9 | 54.5, 57.3 | 53.2, 58.8 | 55.9 | 54.5, 57.3 | 53.2, 58.8 | 55.6 | 54.0, 57.1 | 52.6, 58.6 |
| $q \equiv M_{\mathrm{C}}/M_{\mathrm{B}}$ | 0.951 | 0.918, 0.974 | 0.901, 1.007 | 0.951 | 0.922, 0.978 | 0.895, 1.000 | 0.94 | 0.91, 0.98 | 0.87, 1.00 |
| $T_{\mathrm{eff,B}}$ (K) | 1950 | 1910, 1990 | 1880, 2020 | 1950 | 1910, 1990 | 1880, 2020 | 2000 | 1960, 2030 | 1930, 2070 |
| $T_{\mathrm{eff,C}}$ (K) | 1860 | 1830, 1900 | 1800, 1930 | 1860 | 1830, 1900 | 1790, 1930 | 1920 | 1880, 1950 | 1850, 1980 |
| $\Delta T_{\mathrm{eff}}$ (K) | 90 | 40, 140 | −10, 180 | 90 | 40, 140 | −10, 180 | 80 | 40, 140 | −20, 170 |
| $\log(g_{\mathrm{B}})$ (cgs) | 5.117 | 5.101, 5.133 | 5.085, 5.148 | 5.117 | 5.101, 5.133 | 5.085, 5.148 | 5.160 | 5.144, 5.176 | 5.128, 5.192 |
| $\log(g_{\mathrm{C}})$ (cgs) | 5.095 | 5.079, 5.111 | 5.063, 5.128 | 5.095 | 5.079, 5.111 | 5.063, 5.127 | 5.138 | 5.122, 5.155 | 5.104, 5.172 |
| $R_{\mathrm{B}}$ ($M_{\mathrm{Jup}}$) | 1.054 | 1.045, 1.062 | 1.038, 1.072 | 1.054 | 1.045, 1.062 | 1.038, 1.072 | 1.006 | 0.996, 1.015 | 0.986, 1.026 |
| $R_{\mathrm{C}}$ ($M_{\mathrm{Jup}}$) | 1.055 | 1.046, 1.063 | 1.038, 1.072 | 1.055 | 1.046, 1.063 | 1.038, 1.072 | 1.000 | 0.990, 1.009 | 0.982, 1.019 |
| $(Y − J)_{\mathrm{B}}$ (mag) | 1.135 | 1.117, 1.164 | 1.094, 1.167 | 1.122 | 1.119, 1.126 | 1.111, 1.132 | 1.235 | 1.214, 1.255 | 1.194, 1.274 |
| $(Y − J)_{\mathrm{C}}$ (mag) | 1.138 | 1.122, 1.151 | 1.121, 1.167 | 1.137 | 1.127, 1.144 | 1.123, 1.157 | 1.280 | 1.266, 1.297 | 1.247, 1.310 |
| $(J − H)_{\mathrm{B}}$ (mag) | 0.551 | 0.545, 0.559 | 0.533, 0.581 | 0.560 | 0.543, 0.574 | 0.533, 0.588 | 0.427 | 0.418, 0.435 | 0.410, 0.443 |
| $(J − H)_{\mathrm{C}}$ (mag) | 0.600 | 0.571, 0.633 | 0.550, 0.649 | 0.593 | 0.579, 0.607 | 0.567, 0.630 | 0.405 | 0.395, 0.415 | 0.380, 0.422 |
| $(H − K)_{\mathrm{B}}$ (mag) | 0.464 | 0.447, 0.479 | 0.430, 0.504 | 0.466 | 0.444, 0.487 | 0.428, 0.503 | 0.221 | 0.205, 0.236 | 0.188, 0.248 |
| $(H − K)_{\mathrm{C}}$ (mag) | 0.522 | 0.495, 0.553 | 0.470, 0.574 | 0.511 | 0.494, 0.527 | 0.477, 0.555 | 0.185 | 0.165, 0.203 | 0.142, 0.217 |
| $(J − K)_{\mathrm{B}}$ (mag) | 1.015 | 0.991, 1.037 | 0.962, 1.084 | 1.03 | 0.99, 1.06 | 0.96, 1.09 | 0.648 | 0.622, 0.669 | 0.598, 0.691 |
| $(J − K)_{\mathrm{C}}$ (mag) | 1.12 | 1.07, 1.19 | 1.02, 1.22 | 1.104 | 1.073, 1.133 | 1.045, 1.186 | 0.590 | 0.559, 0.616 | 0.524, 0.639 |
| $(K − L')_{\mathrm{B}}$ (mag) | 0.829 | 0.815, 0.861 | 0.792, 0.861 | 0.822 | 0.808, 0.836 | 0.796, 0.849 | 1.011 | 0.987, 1.035 | 0.962, 1.059 |
| $(K − L')_{\mathrm{C}}$ (mag) | 0.8554 | 0.8510, 0.8580 | 0.8456, 0.8635 | 0.859 | 0.841, 0.873 | 0.829, 0.895 | 1.064 | 1.047, 1.086 | 1.025, 1.105 |

Note. — Saumon & Marley (2008) hybrid and cloudy models give virtually identical physical parameters because the only difference is assuming $f_{sed} = 2$ for the fully cloudy models and $f_{sed} = 1$ for the cloudy portion of the hybrid models.



\end{document}